\documentclass[journal]{IEEEtran}

\usepackage{amsmath}
\usepackage{amssymb}
\usepackage{amsthm}
\usepackage{tikz}
\usetikzlibrary{positioning,fit,arrows.meta,calc}

\newtheorem{lemma}{Lemma}
\newtheorem{theorem}{Theorem}
\newtheorem{definition}{Definition}

\newtheorem{corollary}{Corollary}
\newtheorem{convention}{Convention}

\newcommand{\Btp}{\operatorname{Btp}}
\newcommand{\Bin}{\operatorname{Bin}}
\newcommand{\Pois}{\operatorname{Pois}}
\newcommand{\Ber}{\operatorname{Ber}}
\newcommand{\Del}{\operatorname{Del}}
\newcommand{\BinCh}{\operatorname{BinCh}}
\newcommand{\Var}{\operatorname{Var}}
\newcommand{\BSC}{\operatorname{BSC}}

\begin{document}

\title{Lower Bounds on Deletion Channel Capacity in the High-Deletion Regime via Channel Composition}

\author{
\IEEEauthorblockN{Adel Aghajan and Tolga M. Duman}
\thanks{This work was funded by the European Union
through the ERC Advanced Grant 101054904: TRANCIDS.
Views and opinions expressed are, however, those of
the authors only and do not necessarily reflect those
of the European Union or the European Research Council
Executive Agency. Neither the European Union nor the
granting authority can be held responsible for them.

The authors are with the Electrical and Electronics Engineering Department,
Bilkent University, 06800 Ankara, Turkiye (e-mail: adel.a@bilkent.edu.tr; 
duman@ee.bilkent.edu.tr).}
}

\maketitle

\begin{abstract}
We study repetition-channel capacity through the viewpoint of channel
composition, with particular focus on deletion, binomial-repetition,
and Poisson-repetition channels. This framework can be viewed as a
generalization of the approach introduced by Mitzenmacher and Drinea,
who established a capacity inequality relating the deletion and
Poisson-repetition channels. We develop this viewpoint systematically
and derive several new capacity inequalities relating repetition
channels. In particular, we establish a capacity inequality between
binomial-repetition and Poisson-repetition channels, which leads to
improved lower bounds on the deletion-channel capacity, particularly
in the high-deletion regime. We also characterize the asymptotic
relationship between the normalized deletion and Poisson-repetition
channel capacities, showing that they coincide as the deletion
probability approaches one and the Poisson parameter approaches zero.
Furthermore, we extend our approach to deletion channels with
substitution errors and derive improved capacity lower bounds.
\end{abstract}

\begin{IEEEkeywords}
Repetition channels, synchronization errors, capacity inequality, deletion channels with substitution errors,
Mrs.\ Gerber's lemma.
\end{IEEEkeywords}

\section{Introduction}

Channels with synchronization errors form an important class of
communication models in which the correspondence between transmitted
and received symbol positions is lost. A classical example is the
binary deletion channel, where each transmitted bit is independently
deleted with probability $d$, while the surviving bits are received
without error and in their original order. Although the model is simple
to describe, determining its capacity remains a difficult problem.
General coding theorems for channels with memoryless synchronization errors were
established in~\cite{dobrushin1967shannon}.

A major line of work on lower bounding the capacity of the deletion
channel is based on the information-theoretic characterization of
synchronization-channel capacity in~\cite{dobrushin1967shannon}. In
particular, for the deletion channel, the result in
\cite{dobrushin1967shannon} gives
\begin{align}\label{eqn:dobrushineqn}
C(\Del(d))
=
\lim_{n\to\infty}
\frac{1}{n}
\max_{P_{X^n}} I(X^n;Y),
\end{align}
where $C(\Del(d))$ denotes the capacity of the deletion channel with
deletion probability $d$ and $Y$ denotes the output of the deletion channel corresponding to
the input $X^n$. Consequently, by choosing a particular input process
and lower bounding its asymptotic mutual-information rate, one obtains
a lower bound on the deletion-channel capacity.

This viewpoint has led to several important lower bounds. Early works
considered uniformly distributed inputs and low-order Markov input
processes. Analytical lower bounds based on first-order Markov inputs
were derived in~\cite{diggavi2006information}, while improved bounds
based on run-length representations and jigsaw decoding were developed
in~\cite{drinea2006lower,drinea2007improved}. A more direct use of the
characterization in \eqref{eqn:dobrushineqn} was considered in
\cite{kirsch2010directly}, where the asymptotic mutual-information rate
was lower bounded for input processes with general independent and identically distributed (i.i.d.)\ run lengths.
Computable lower bounds based on a decomposition of the
mutual-information rate for first-order Markov inputs were obtained in
\cite{venkataramanan2013achievable}. Numerical and simulation-based
methods have also been used to estimate achievable information rates
for deletion and related synchronization error channels, using both i.i.d.\
and Markov input sources
\cite{vvedenskaya1968computation,kavcic2004insertion,
castiglione2015trellis,hu2010achievable,mercier2012bounds}. A detailed
overview of these approaches can be found in
\cite{cheraghchi2021overview}.

Recent preprints have reported further progress on both
upper and lower bounds for the binary deletion channel.
Using an AI-assisted approach,
\cite{soysal2026upper} established the high-deletion
upper bound $C(\Del(d))\leq(1-d)/4$ for $d\geq0.65$,
with a finite certificate verified using exact arithmetic
and a proof formalized in Lean.
Similarly, \cite{papailiopoulos2026approximation}
employed interacting language-model agents to develop
computer-assisted lower and upper bounds whose midpoint
approximates the channel capacity with an error of less
than $0.0095$ bits per transmitted bit, uniformly over
all deletion probabilities.
On the achievability side,
\cite{khodaiemehr2026learning} employed a learning-based
approach to optimize input run-length distributions,
obtaining improved lower bounds certified using
interval arithmetic.
The lower bounds in \cite{papailiopoulos2026approximation}
are generally tighter for intermediate deletion probabilities,
whereas \cite{khodaiemehr2026learning} provides stronger
lower bounds at certain low and high deletion probabilities,
particularly at $d=0.01$, $d=0.80$, and $d=0.90$.

A common feature of these direct approaches is that
the information rate is evaluated or bounded at
specific deletion probabilities. The monotonicity
of the normalized deletion-channel capacity,
established in \cite{rahmati2014upper},
allows a lower bound obtained at a deletion probability
$d$ to be extended to smaller deletion probabilities.
However, this property does not allow the lower bound
to be extended to deletion probabilities greater than $d$.
This limitation is particularly relevant in the
high-deletion regime, where the channel capacity
approaches zero as $d\to1$.


The high-deletion regime is particularly challenging.
The classical lower bound
$C(\Del(d))\geq 1-h_2(d)$, valid for
$0\leq d\leq 1/2$, becomes trivial at $d=1/2$
and does not provide a useful bound in the
high-deletion regime. Obtaining stronger bounds
typically requires input processes with memory
\cite{diggavi2006information}.
When $d$ is close to one, only a small fraction
of the transmitted symbols survive, and their
original positions are unknown to the receiver.
This loss of synchronization makes direct
information-rate calculations particularly difficult.
Moreover, the computational complexity generally
increases with the blocklength and the complexity
of the input process. For example, finite-block
mutual-information optimization can, in principle,
be performed numerically using the
Blahut--Arimoto algorithm, but this approach
quickly becomes computationally infeasible as
the blocklength grows \cite{fertonani2010novel}.
Consequently, although the deletion-channel
capacity is understood more precisely in the
low-deletion regime, a substantial gap between
the known normalized upper and lower bounds
remains in the high-deletion regime
\cite{cheraghchi2021overview}.

A different approach was introduced in
\cite{mitzenmacher2006simple}. Rather than analyzing the deletion channel
directly for each value of $d$, the Poisson-repetition
channel was introduced and a coding-theoretic connection between the
Poisson-repetition and deletion channels was established. The
construction implies that, for every $\lambda>0$ and every $0\leq d<1$, we have
\begin{align*}
\frac{C(\Del(d))}{1-d}
\geq
\frac{C(\Pois(\lambda))}{\lambda},
\end{align*}
where $C(\Pois(\lambda))$ denotes the capacity of the Poisson-repetition
channel with mean $\lambda$.
An important feature of this relation is that a lower bound on the
capacity of a \emph{single} Poisson-repetition channel immediately
gives a lower bound on the deletion-channel capacity for every
deletion probability $d$. Thus, unlike a direct information-rate
calculation for the deletion channel, the Poisson channel needs to be
analyzed only once.

Using this relation together with a lower bound on the capacity of the
Poisson-repetition channel, \cite{mitzenmacher2006simple} obtained
\begin{align*}
C(\Del(d))
\geq
0.1185(1-d),
\qquad 0\leq d<1.
\end{align*}
Although this bound holds for all deletion probabilities, it is
particularly useful in the high-deletion regime, where several other
lower bounds become weak or trivial. More recently,
\cite{rubinstein2023improved} considered a broader
class of input run-length distributions for the Poisson-repetition
channel and improved the coefficient from $0.1185$ to $0.1221$.

The construction of~\cite{mitzenmacher2006simple} can therefore be
viewed as fundamentally different from directly evaluating the
mutual-information rate of the deletion channel. Instead, one first
constructs a code for an auxiliary repetition channel and then
transforms it into a code for the deletion channel. In this paper, we
develop this coding-based viewpoint further and interpret it as a
general composition principle for repetition channels. By inserting
suitable repetition stages {\it before} or {\it after} a given
repetition channel, we obtain new capacity comparisons between
deletion, binomial-repetition, and Poisson-repetition channels.

Among the consequences of this framework, we highlight two main
results. First, our finite-deletion result shows that a suitable
binomial-repetition construction yields lower bounds on
$C(\Del(d))$ that are strictly stronger than the classical
Poisson bound in~\cite{mitzenmacher2006simple} for a range of deletion
probabilities $d<1$ in the high-deletion regime.

Second, our asymptotic result characterizes precisely the connection
between the deletion and Poisson-repetition channels. We prove that
\begin{align*}
\lim_{d\to1}
\frac{C(\Del(d))}{1-d}
=
\lim_{\lambda\to0}
\frac{C(\Pois(\lambda))}{\lambda}.
\end{align*}
Consequently, the Poisson-based lower bound becomes asymptotically
tight when the Poisson parameter is allowed to approach zero. This
resolves the asymptotic tightness question raised
in~\cite{mitzenmacher2006simple} and gives a precise
characterization of the relation between the two channels in the
high-deletion limit.

We also extend our approach to the concatenation of a
deletion channel and a binary symmetric channel (BSC),
which incorporates both synchronization and substitution
errors. Capacity bounds for channels with both types
of errors were first investigated by Gallager
\cite{gallager1961sequential}, and subsequently extended
to deletion channels followed by general discrete
memoryless channels in \cite{diggavi2006information}.
Further upper and lower bounds were developed in
\cite{fertonani2011bounds,rahmati2014achievable}.
In particular, \cite{rahmati2014achievable} derived
lower bounds for the deletion--BSC channel by relating
its capacity to that of the underlying deletion channel.
Building on these results, we extend our channel-composition
framework to the deletion--BSC channel and employ
Mrs.\ Gerber's Lemma to obtain improved capacity lower
bounds. Specifically, we use symmetric first-order Markov
inputs to bound the conditional entropy of the
repetition-channel output, leading to explicit lower
bounds for the deletion--BSC channel, including in the
high-deletion regime.

The remainder of the paper is organized as follows.
Section~\ref{sec:main_results} presents the main results
and their consequences for deletion, binomial-repetition,
and Poisson-repetition channels.
Section~\ref{sec:del_bsc} extends the framework to
deletion channels followed by a binary symmetric channel
and derives improved capacity lower bounds.
Sections~\ref{sec:proof_del} and
\ref{sec:proof_del_BSC} provide the proofs of the
deletion channel and deletion--BSC results, respectively.
Finally, Section~\ref{sec:Conclusion} concludes the paper.



\section{Deletion Channels}
\label{sec:main_results}

The central idea of this paper is to use composition of repetition
channels as a tool for comparing their capacities. By inserting a
suitable random repetition stage before or after a given repetition
channel, one can transform it into another repetition channel whose
capacity is better understood. This simple viewpoint leads to several
consequences for deletion, binomial repetition, and Poisson repetition
channels. The main results are presented in this section, and their proofs are provided in the following section.

We begin by introducing the class of repetition channels considered
throughout the paper. This class contains the deletion, Poisson
repetition, and binomial repetition channels as special cases.

\begin{definition}[$Q$-Repetition Channel]
\label{def:q_repeat}
Let $Q$ be a probability distribution on the nonnegative integers. A
$Q$-repetition channel is a synchronization channel in which each input
symbol is independently repeated a $Q$-distributed number of times.
More precisely, consider an input sequence
\begin{align*}
X=X_1X_2\cdots X_n,
\end{align*}
and let $N_1,\ldots,N_n$ be i.i.d.\ random variables with $N_i\sim Q$.
The output of the channel is
\begin{align*}
Y=X_1^{N_1}X_2^{N_2}\cdots X_n^{N_n},
\end{align*}
where $x^N$ denotes $N$
consecutive copies of $x$, with $x^0$ denoting the empty string.
\end{definition}

Three choices of $Q$ are of particular interest in this paper. If
$Q=\Ber(1-d)$, each input symbol is retained once with probability $1-d$
and deleted with probability $d$. The corresponding $Q$-repetition
channel is therefore the i.i.d.\ deletion channel, denoted by
$\Del(d)$. We also consider the Poisson repetition channel, corresponding
to $Q=\Pois(\lambda)$, and the binomial repetition channel, corresponding
to $Q=\Bin(m,p)$. 
\begin{convention}
    For a $Q$-repetition channel, we denote its capacity by $C(Q)$. Also, $C(\Del(d))$ denote the capacity of the deletion channel $\Del(d)$.
\end{convention}

\subsection{Repetition before the Channel}

\begin{figure}[t]
    \centering
    \resizebox{0.85\columnwidth}{!}{%
    \begin{tikzpicture}[
        >=stealth,
        node distance=1.8cm,
        block/.style={
            draw,
            rectangle,
            minimum width=2.4cm,
            minimum height=1.4cm,
            align=center
        }
    ]

    \node[block] (rep) {Repetition};
    \node[block, right=of rep] (ch) {Original\\Channel};

    \draw[->] ([xshift=-1.5cm]rep.west) -- (rep.west);
    \draw[->] (rep.east) -- (ch.west);
    \draw[->] (ch.east) -- ([xshift=1.5cm]ch.east);

    \node[
        draw,
        rounded corners,
        fit=(rep)(ch),
        inner xsep=0.7cm,
        inner ysep=0.8cm,
        label=above:{New Channel}
    ] {};

    \end{tikzpicture}%
    }
    \caption{Repetition before the original channel.}
    \label{fig:pre_channel_repetition}
\end{figure}
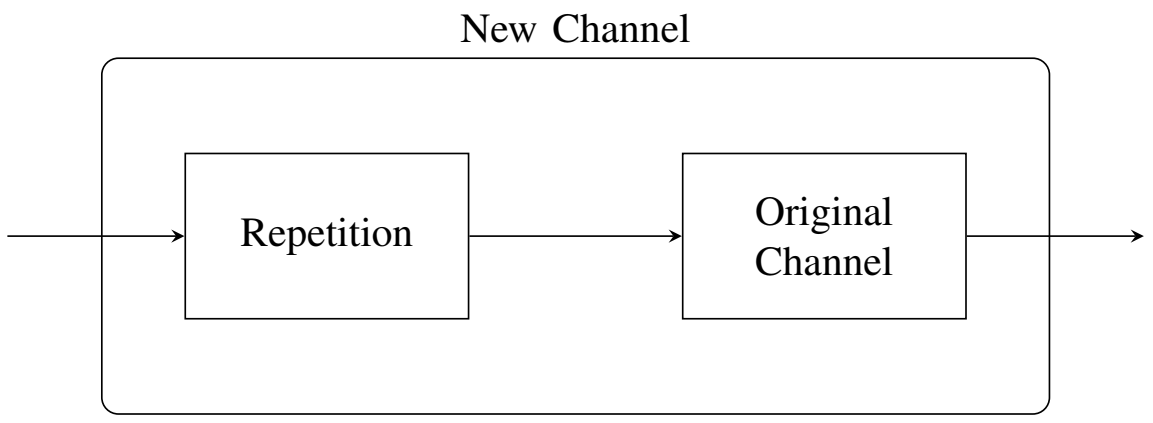

The first comparison mechanism considered in this section is based on
 repeating the input symbols  a random number of times before passing them through a
repetition channel. 
This construction is illustrated in
Fig.~\ref{fig:pre_channel_repetition}.
This repetition stage can be viewed as part of the
encoding process for the original channel. If the resulting cascade is equivalent to another repetition channel, then any code for the latter channel can be converted into a code for the former, with the rate loss determined by the average increase in the encoder blocklength. This idea was first used in \cite{mitzenmacher2006simple} to relate the deletion and Poisson repetition channels. The following lemma formalizes the same idea for general repetition distributions.

\begin{lemma}\label{lem:MainPostEncoding}
Let $P$, $Q$, and $Q'$ be probability distributions on
$\mathbb Z_{\geq 0}$. Let $A\sim P$, and, independently of $A$, let
$B_1,B_2,\ldots$ be i.i.d. random variables with distribution $Q$.
Suppose that
\begin{align*}
\sum_{j=1}^{A} B_j \sim Q',
\end{align*}
where the sum is defined to be zero when $A=0$. 
Assume that $0<\mathbb{E}[A]<\infty$ and
$\Var(A)<\infty$. Then
\begin{align*}
C(Q)\geq \frac{1}{\mathbb{E}[A]}C(Q').
\end{align*}
\end{lemma}

Lemma~\ref{lem:MainPostEncoding} provides a general tool for transferring
capacity lower bounds between repetition channels. Several useful
relations follow by choosing the two repetition distributions
appropriately. We first recall the corresponding Poisson construction
for the deletion channel and then derive its binomial analogue.

In \cite{mitzenmacher2006simple}, the choice
$P=\Pois\!\left(\frac{\lambda}{1-d}\right)$,
$Q=\Ber(1-d)$, and $Q'=\Pois(\lambda)$ yields, for every
$0\leq d\leq1$ and every $\lambda>0$,
\begin{align}\label{eq:PoisLowerDel}
\frac{C(\Del(d))}{1-d}
\geq
\frac{C(\Pois(\lambda))}{\lambda}.
\end{align}
This relation is particularly useful in the high-deletion regime
$d\to1$.

A similar construction is possible for the binomial repetition channel.
Specifically, taking
$P=\Bin\!\left(m,\frac{p}{1-d}\right)$,
$Q=\Ber(1-d)$, and $Q'=\Bin(m,p)$
gives the following binomial counterpart of
\eqref{eq:PoisLowerDel}.

\begin{corollary}[Binomial Lower Bound]
\label{thm:bin_lower_del}
For every $m\geq1$, $p\in[0,1]$, and
$0\leq d\leq1-p,$
we have
\begin{align}
\label{eq:BinLowerDel}
\frac{C(\Del(d))}{1-d}
\geq
\frac{C(\Bin(m,p))}{mp}.
\end{align}
\end{corollary}

Unlike the Poisson bound \eqref{eq:PoisLowerDel},
Corollary~\ref{thm:bin_lower_del} does not apply to every $d$ for a fixed
$p$. It requires $d\leq1-p$. Nevertheless, by choosing $p$ sufficiently
small, the bound can be applied arbitrarily close to $d=1$. Moreover, if
$p\to0$ and $m\to\infty$ while $mp=\lambda$ is kept fixed, then
$\Bin(m,p)$ converges to $\Pois(\lambda)$. Consequently, in the limiting
high-deletion regime, the Binomial and Poisson approaches recover the
same lower bound for deletion channels.

Since the deletion channel $\Del(d)$ is equivalent to the binomial
repetition channel $\Bin(1,1-d)$, Corollary~\ref{thm:bin_lower_del}
 gives the following result.

\begin{corollary}[Monotonicity of Normalized Deletion Capacity]\label{cor:CDelnonincreasing}
$C(\Del(d))/(1-d)$ is nonincreasing in $d$.
\end{corollary}

Corollary~\ref{cor:CDelnonincreasing} can also be obtained from
\cite[Theorem~1]{rahmati2014upper}. 
Another consequence of Lemma~\ref{lem:MainPostEncoding} is obtained by
using deterministic repetition before a Poisson repetition channel.
Repeating each input symbol $m$ times before a $\Pois(\lambda)$
repetition channel produces a $\Pois(m\lambda)$ repetition channel.
Lemma~\ref{lem:MainPostEncoding} therefore gives the following scaling
relation.

\begin{corollary}[Poisson Scaling Relation]
\label{cor:PoisChScaling}
For every integer $m\geq1$ and every $\lambda>0$, we have
\begin{align}
\label{eq:PoisChDecreasing}
C(\Pois(\lambda))
\geq
\frac{C(\Pois(m\lambda))}{m}.
\end{align}
\end{corollary}

Equivalently, Corollary~\ref{cor:PoisChScaling} implies
\begin{align*}
\frac{C(\Pois(\lambda))}{\lambda}
\geq
\frac{C(\Pois(m\lambda))}{m\lambda}.
\end{align*}
This scaling relation suggests a possible monotonic behavior of the
normalized quantity $C(\Pois(\lambda))/\lambda$ as $\lambda$ varies,
and motivates the question of whether its limit exists as
$\lambda\to0$. We return to this question in the last subsection.

\subsection{Repetition after the Channel: Data-Processing Inequality}

\begin{figure}[t]
    \centering
    \resizebox{0.85\columnwidth}{!}{%
    \begin{tikzpicture}[
        >=stealth,
        node distance=1.8cm,
        block/.style={
            draw,
            rectangle,
            minimum width=2.4cm,
            minimum height=1.4cm,
            align=center
        }
    ]

    \node[block] (ch) {Original\\Channel};
    \node[block, right=of ch] (rep) {Repetition};

    \draw[->] ([xshift=-1.5cm]ch.west) -- (ch.west);
    \draw[->] (ch.east) -- (rep.west);
    \draw[->] (rep.east) -- ([xshift=1.5cm]rep.east);

    \node[
        draw,
        rounded corners,
        fit=(ch)(rep),
        inner xsep=0.7cm,
        inner ysep=0.8cm,
        label=above:{New Channel}
    ] {};

    \end{tikzpicture}%
    }
    \caption{Repetition after the original channel.}
    \label{fig:post_channel_repetition}
\end{figure}
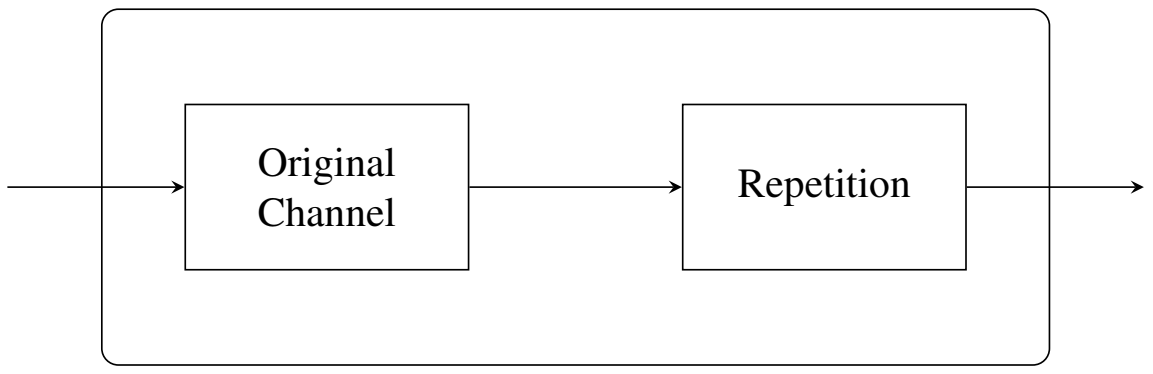

The preceding results compare repetition channels by modifying the
encoder before transmission. In this section, we instead introduce
a repetition stage after the original channel and consider the case
in which the resulting output has the same distribution as that of
another repetition channel. This construction is illustrated in
Fig.~\ref{fig:post_channel_repetition}. Unlike repetition before
the channel, which increases the number of transmitted symbols,
repetition after the channel can be viewed as part of the decoding
process, since it operates only on the received sequence.
Consequently, the input blocklength remains unchanged, and no
normalization by the expected repetition factor is required.
By the data-processing inequality, the capacity of the original
channel is therefore at least as large as that of the resulting
channel.

We saw that \eqref{eq:PoisLowerDel} and \eqref{eq:BinLowerDel} provide
lower bounds on the deletion channel capacity. A natural question is
which of these two constructions  yields a stronger lower bound.
To address this question, we introduce an additional repetition stage
after the binomial repetition channel and obtain the following theorem.

\begin{theorem}[Binomial--Poisson Comparison]
\label{thm:bin_pois_comparison}
Let $Y$ and $Z$ denote the outputs of a $\Bin(m,p)$ repetition channel
and a $\Pois(\lambda)$ repetition channel, respectively, driven by the
same input sequence $X$. Suppose that
\begin{align*}
p\geq 1-e^{-\lambda/m}.
\end{align*}
Then, for every distribution on $X$, $I(X;Y)\geq I(X;Z)$
Consequently, we have $C\bigl(\Bin(m,p)\bigr)
\geq
C\bigl(\Pois(\lambda)\bigr).$
\end{theorem}

Theorem~\ref{thm:bin_pois_comparison} can now be combined  with
Corollary \ref{thm:bin_lower_del} and  gives the following corollary.

\begin{corollary}[Deletion-Channel Lower Bound]
\label{cor:del_lower_bound}
Suppose $p\geq 1-e^{-\lambda/m}$ and $d\leq 1-p$. For the i.i.d.\
deletion channel with deletion probability $d$, we have
\begin{align}
\label{eq:DelLowerFromPois}
\frac{C(\Del(d))}{1-d}
\geq
\frac{C(\Pois(\lambda))}{mp}.
\end{align}
\end{corollary}

Corollary~\ref{cor:del_lower_bound} allows us to compare the new
construction directly with the classical Poisson lower bound. Two
choices of the binomial parameter $p$ are particularly informative.
The first choice is $p=\lambda/m$ and the second choice is $p\geq 1-e^{-\lambda/m}$.

If $p=\lambda/m$, then $p\geq 1-e^{-\lambda/m}$
which follows from $e^{-x}\geq1-x$ for $x\geq0$.
Therefore, for $\lambda=mp$, we have
\begin{align*}
C\bigl(\Bin(m,p)\bigr)
\geq
C\bigl(\Pois(mp)\bigr).
\end{align*}
In particular, when the two repetition distributions have the same
mean, the Binomial-repetition channel is never worse than the Poisson-repetition channel in
terms of capacity. Therefore, Corollary~\ref{cor:del_lower_bound} shows
that the Binomial construction does not yield a worse deletion-channel
lower bound than the corresponding Poisson construction whenever the
Binomial bound is applicable.


However, the Binomial construction can in fact yield a strictly better
lower bound by choosing $p=1-e^{-\lambda/m}$. This is becuase for every $\lambda>0$ and finite $m$, we have
$1-e^{-\lambda/m}<\frac{\lambda}{m}.$
Hence, choosing $p=1-e^{-\lambda/m}$
gives $mp<\lambda$. Corollary~\ref{cor:del_lower_bound} therefore yields,
for every $d\leq e^{-\lambda/m}$,
\begin{align*}
\frac{C(\Del(d))}{1-d}
\geq
\frac{C(\Pois(\lambda))}
{m(1-e^{-\lambda/m})}
>
\frac{C(\Pois(\lambda))}{\lambda}.
\end{align*}

Thus, for every finite $m$, the Binomial construction gives a strict
improvement over the Poisson construction whenever
$d\leq e^{-\lambda/m}$.
In particular, at the boundary value $d=e^{-\lambda/m}$, we obtain the
simple bound
\begin{align*}
C\bigl(\Del(e^{-\lambda/m})\bigr)
\geq
\frac{1}{m}C\bigl(\Pois(\lambda)\bigr).
\end{align*}


The improvement depends on the integer parameter $m$.
For a fixed $\lambda$, since $m(1-e^{-\lambda/m})$ is an increasing
function of $m$, the lower bound on $C(\Del(d))/(1-d)$ is maximized by
choosing the smallest feasible integer $m$. Since the bound requires
$d\leq e^{-\lambda/m}$, we must have
$m\geq \lambda/(-\log d)$. Therefore, the smallest feasible integer is
\begin{align*}
m^*(d)=\left\lceil \frac{\lambda}{-\log d}\right\rceil.
\end{align*}
Moreover, for
$d\in\left(e^{-\lambda/(m-1)},e^{-\lambda/m}\right]$, we have
$m^*(d)=m$. Hence, over this interval, the resulting lower bound is
\begin{align*}
\frac{C(\Del(d))}{1-d}
\geq
\frac{C(\Pois(\lambda))}
{m(1-e^{-\lambda/m})}.
\end{align*}

The preceding expressions can be evaluated using any available lower
bound on the capacity of a Poisson-repetition channel. To illustrate
the resulting tradeoff, we use two previously reported Poisson-repetition channel
bounds. One has a larger asymptotic coefficient, whereas the other can
be applied over a substantially wider range of deletion probabilities.

\begin{figure}
    \centering
    \includegraphics[width=1\linewidth]{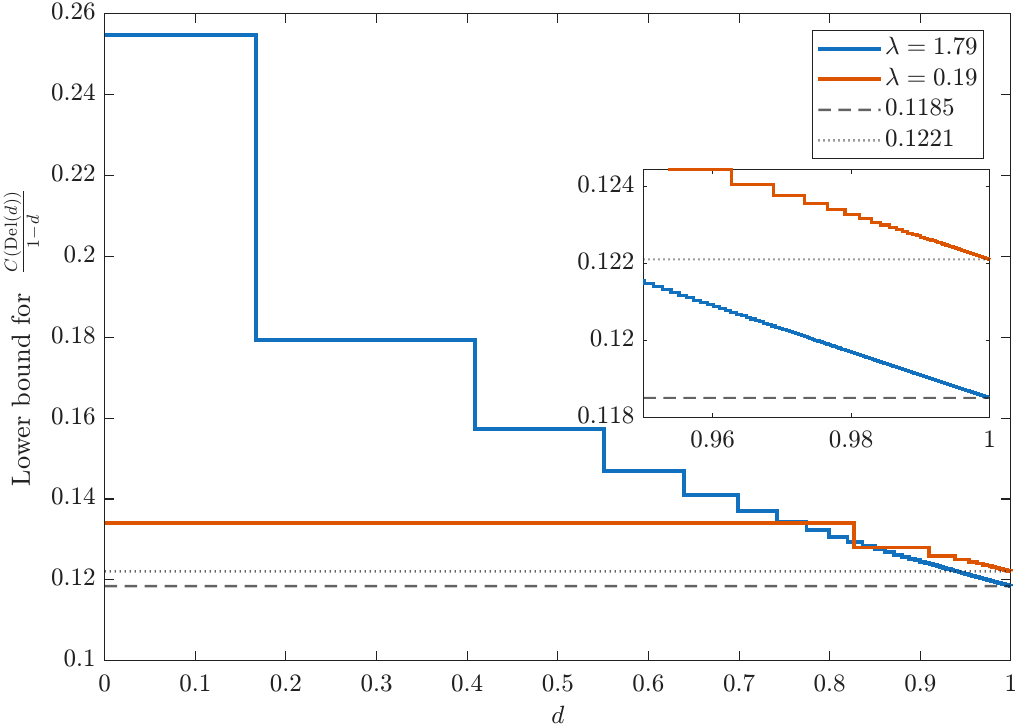}
    \caption{Normalized deletion-channel lower bound obtained using
    $\lambda=1.79$ and $\lambda=0.19$}
    \label{fig:lambda179}
\end{figure}

In \cite{mitzenmacher2006simple}, the lower bound
$C(\Pois(1.79))\geq 1.79\times 0.1185$ is reported, while in
\cite{rubinstein2023improved}, the lower bound
$C(\Pois(0.19))\geq 0.19\times 0.1221$ is reported. In
Fig.~\ref{fig:lambda179}, we plot the resulting
normalized deletion channel lower bound
\[C(\Pois(\lambda))/(m(1-e^{-\lambda/m}))\] 
for these two choices of
$\lambda$.

The choice $\lambda=0.19$ provides the larger asymptotic coefficient,
namely $0.1221$ instead of $0.1185$. However, the bound corresponding
to $m=1$ starts at $d=e^{-0.19}$, so this choice is useful for larger
deletion probabilities. In contrast, for $\lambda=1.79$, the first
interval starts at $d=e^{-1.79}$. Consequently, although its asymptotic
coefficient is smaller, the choice $\lambda=1.79$ provides an improved
bound over a wider range of deletion probabilities.

So far, we have used Poisson-repetition channel lower bounds to obtain improved
lower bounds for the capacity of  deletion channels. The same comparison can also be
used in the reverse direction. Namely, an upper bound on the normalized
deletion channel capacity over a sufficiently large range of deletion
probabilities yields an upper bound on the capacity of a Poisson
repetition channel. Optimizing over the admissible integer $m$ gives
the following result.

\begin{corollary}\label{cor:UpperBoundPoissonCh}
Let $C(\Del(d))\le \alpha(1-d)$ for all
$d\ge \beta$.
Then, we have
\begin{align*}
C(\operatorname{Pois}(\lambda))
\le
\alpha\,m_\lambda(1-e^{-\lambda/m_\lambda}),
\end{align*}
where
$m_\lambda=
\max\left\{1,
\left\lceil\frac{\lambda}{-\log(\beta)}\right\rceil\right\}.$
\end{corollary}

In \cite{rahmati2014upper}, upper bounds of the form
$C(\Del(d))\leq\alpha(1-d)$ for $d\geq\beta$
were obtained. Following a similar approach,
these bounds were subsequently improved through
more efficient implementations of the
Blahut--Arimoto algorithm and more extensive
numerical computations
\cite{rubinstein2023improved,pinto2026improved}.

 The best currently known capacity upper bound of the above form is given in \cite{soysal2026upper},
which establishes
\begin{align*}
C(\Del(d))
\leq
(1-d)/4,
\qquad d\geq0.65.
\end{align*}
Therefore, Corollary~\ref{cor:UpperBoundPoissonCh} gives
\begin{align*}
C(\operatorname{Pois}(\lambda))
\leq
\frac{m_\lambda}{4}
\left(1-e^{-\lambda/m_\lambda}\right),
\end{align*}
where $m_\lambda=\max\left\{1,\left\lceil\frac{\lambda}{-\log(0.65)}\right\rceil\right\}.$
This improves the upper bound in
\cite{cheraghchi2019capacity}, which gives
\begin{align*}
C(\operatorname{Pois}(\lambda))
\leq
0.464421\left(1-e^{-\lambda}\right)(1+o(1))
\end{align*}
for small $\lambda$. 

\subsection{Repetition after the Channel: Beyond Data-Processing Inequality}
The Poisson construction in \eqref{eq:PoisLowerDel}, obtained in
\cite{mitzenmacher2006simple}, provides one direction of comparison
between the normalized capacities of the deletion and Poisson-repetition
channels. In \cite{mitzenmacher2006simple}, the authors also ask whether
this bound is asymptotically tight, namely, whether there exists a
constant $\lambda>0$ such that
\begin{align*}
\lim_{d\to1}
\frac{C(\Del(d))}{1-d}
=
\frac{C(\Pois(\lambda))}{\lambda}.
\end{align*}

In this section, we study a more general asymptotic question. Since the
deletion channel is the special case  of a binomial
repetition channel, we consider, for every fixed integer $m\geq1$, the
relationship between the normalized capacities of the
$\Bin(m,p)$ and Poisson-repetition channels as $p\to0$. We show that these normalized capacities have the same
limit. The deletion-channel result then follows immediately by taking
$m=1$.

To establish this result, we again introduce an additional repetition
stage after the original channel. In contrast to the previous capacity
comparisons, however, the data-processing inequality alone is not
sufficient, since here we seek an inequality in the opposite direction.
We want a lower bound on the capacity of the resulting composite channel
in terms of the capacity of the original channel.
To address this question, we first consider the quantity
\begin{align*}
\sup_{\lambda>0}
\frac{C(\Pois(\lambda))}{\lambda},
\end{align*}
which is characterized in the following lemma.

\begin{lemma}[Limit of the Normalized Poisson Capacity]
\label{lem:LimitPoissonOverLambda}
The limit of $
{C(\Pois(\lambda))}/{\lambda}$ as $\lambda\to 0$
exists, and
\begin{align*}
\lim_{\lambda\to 0}
\frac{C(\Pois(\lambda))}{\lambda}
=
\sup_{\lambda>0}
\frac{C(\Pois(\lambda))}{\lambda}.
\end{align*}
\end{lemma}

Lemma~\ref{lem:LimitPoissonOverLambda} shows that the largest normalized
Poisson capacity is attained in the limit as $\lambda\to0$. We now
compare this limiting quantity with the normalized capacity of the
$\Bin(m,p)$ repetition channel as $p\to0$. The next theorem shows that,
for every fixed integer $m\geq1$, these two limiting normalized
capacities are equal.

\begin{theorem}[Asymptotic Equivalence of Binomial and Poisson Capacities]
\label{thm:BinPoisAsymptotic}
For every fixed integer $m\geq1$, we have
\begin{align*}
\lim_{p\to0}
\frac{C(\Bin(m,p))}{mp}
=
\lim_{\lambda\to0}
\frac{C(\Pois(\lambda))}{\lambda}.
\end{align*}
\end{theorem}

Since the deletion channel $\Del(d)$ is equivalent to the
$\Bin(1,1-d)$ repetition channel, Theorem~\ref{thm:BinPoisAsymptotic}
immediately yields the following corollary.

\begin{corollary}
\label{cor:DelPoisAsymptotic}
We have
\begin{align*}
\lim_{\lambda\to0}
\frac{C(\Pois(\lambda))}{\lambda}
=
\lim_{d\to1}
\frac{C(\Del(d))}{1-d}.
\end{align*}
\end{corollary}

To prove Theorem~\ref{thm:BinPoisAsymptotic}, we place an additional
repetition channel after the $\Bin(m,p)$ repetition channel so that the
resulting cascade is exactly a Poisson-repetition channel. As the
parameters approach zero, the distribution of this additional
repetition stage converges to the deterministic distribution
concentrated at one. In other words, the additional repetition channel
approaches the identity channel and therefore asymptotically has no
effect on the transmitted sequence. This suggests that, in the
small-parameter regime, the binomial and Poisson repetition channels
should have the same normalized capacity. Making this argument
rigorous requires controlling the information loss introduced by the
additional repetition stage. The technical details are provided in the
next section.
\section{ Deletion Channels  concatenated with  Binary Symmetric Channels}
\label{sec:del_bsc}

In this section, we consider a binary deletion channel
followed by a binary symmetric channel (BSC), denoted by
$\Del(d)+\BSC(\epsilon)$. More generally, we denote
a $Q$-repetition channel 
followed by a BSC with crossover probability $\epsilon$,
by $Q+\BSC(\epsilon)$.
Several lower bounds on the capacity of the
$\Del(d)+\BSC(\epsilon)$ channel have been established in the
literature. Our main objective is to extend the
repetition-based approach developed in the previous
section to this setting and obtain improved capacity
lower bounds, particularly in the high-deletion regime.
Nevertheless, our approach also yields improvements
for other deletion probabilities as well.

To extend the channel-composition approach to this setting, consider a repetition channel with
repetition distribution $Q$, preceded by an independent
repetition process with distribution $P$. The resulting
channel is again a repetition channel, with a new
repetition distribution $Q'$ determined by $P$ and $Q$.
When an independent BSC is applied after the repetition
channel, the same construction remains valid, since
each output bit passes independently through the BSC\footnote{In contrast, repeating bits after the BSC does not
produce the same channel model, since all copies of
a given bit share the same realization of the
substitution noise.}.
Therefore, the capacity comparisons
obtained by repeating bits before a channel extend
directly to the corresponding channels followed by
a BSC. In particular, we obtain the following corollary.

\begin{corollary}
\label{cor:DelPoisBSC}
For every $0\leq d<1$, $\lambda>0$, and
$0\leq\epsilon\leq1/2$, we have
\begin{align}
\frac{C(\Del(d)+\BSC(\epsilon))}{1-d}
\geq
\frac{C(\Pois(\lambda)+\BSC(\epsilon))}{\lambda}.
\label{eq:DelLowerFromPois}
\end{align}
\end{corollary}

In view of Corollary~\ref{cor:DelPoisBSC},
a lower bound on the capacity of the
$\Pois(\lambda)+\BSC(\epsilon)$ channel
immediately yields a lower bound on the capacity
of the $\Del(d)+\BSC(\epsilon)$ channel.
Therefore, we  focus on deriving  lower bounds
for the capacity of ${\Pois(\lambda)+\BSC(\epsilon)}$.

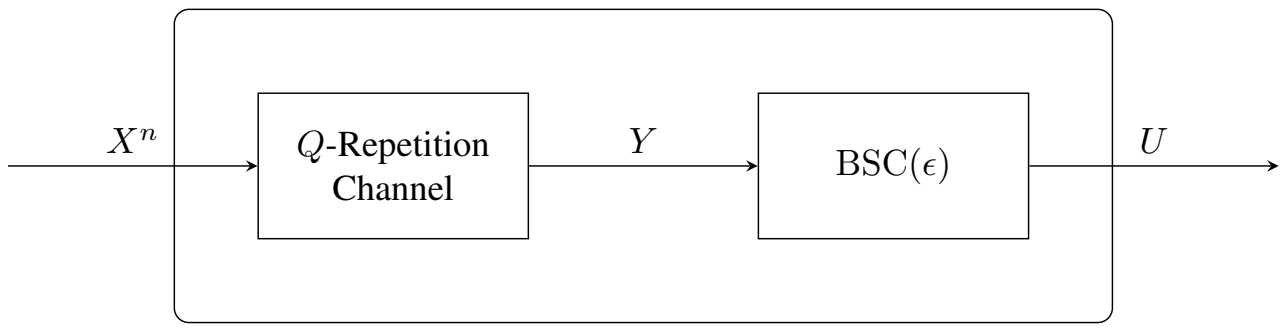
\begin{figure}[t]
    \centering
    \resizebox{0.95\columnwidth}{!}{%
    \begin{tikzpicture}[
        >=stealth,
        node distance=2.2cm,
        block/.style={
            draw,
            rectangle,
            minimum width=2.6cm,
            minimum height=1.4cm,
            align=center
        }
    ]

    \node[block] (rep) {$Q$-Repetition\\Channel};
    \node[block, right=of rep] (bsc) {$\BSC(\epsilon)$};

    \draw[->] ([xshift=-2.4cm]rep.west) -- node[above] {$X^n$} (rep.west);
    \draw[->] (rep.east) -- node[above] {$Y$} (bsc.west);
    \draw[->] (bsc.east) -- node[above] {$U$} ([xshift=2.4cm]bsc.east);

    Outer box
    \node[
        draw,
        rounded corners,
        fit=(rep)(bsc),
        inner xsep=0.8cm,
        inner ysep=0.8cm,
        label=above:{\(Q+\BSC(\epsilon)\) Channel}
    ] {};

    \end{tikzpicture}%
    }
    \caption{The \(Q+\BSC(\epsilon)\) channel. }
    \label{fig:pois_bsc_channel}
\end{figure}

To this end, we build upon the approach of
Rahmati and Duman~\cite{rahmati2014achievable}.
Consider the Markov chain
$X^n\to Y\to U$, where $X^n$ denotes the
channel input, $Y$ is the output of the deletion
channel, and $U$ is obtained by passing $Y$
through an independent $\BSC(\epsilon)$.
This channel cascade is illustrated in
Fig.~\ref{fig:pois_bsc_channel} for the more
general case of a repetition channel followed
by a BSC.

By the chain rule, one can obtain\footnote{We derive \eqref{eq:del_bsc_mutual_information_bound}  in the proof of Lemma \ref{lem:QPlusBSC}.}
\begin{align}
I(X^n;U)
&\geq I(X^n;Y)-H(Y\mid U).
\label{eq:del_bsc_mutual_information_bound}
\end{align}
Moreover, one can write
\begin{align}
H(Y\mid U)
&=H(U\mid Y)+H(Y)-H(U)
\label{eq:del_bsc_entropy_identity}\\
&\leq H(U\mid Y)
\nonumber\\
&=n\lambda h_2(\epsilon).
\label{eq:del_bsc_existing_bound}
\end{align}
where the last equality follows from fact that the BSC acts independently on each
output bit.

Although \eqref{eq:del_bsc_existing_bound}
holds for arbitrary input distributions,
it can be conservative because it does not
exploit the statistical structure of the
repetition-channel output. For example,
in the deletion channel with independent
and uniformly distributed input bits, the
output is uniformly distributed. Therefore,  we have $H(Y)=H(U)$,
and hence
\begin{align}
H(Y\mid U)=H(U\mid Y).
\end{align}
However, this equality does not generally
hold for correlated input distributions or
for repetition channels that can produce
multiple copies of an input bit.

In particular, correlated input distributions,
such as first-order Markov inputs, are important
for synchronization channels, where exploiting
the memory structure of the input can improve
achievable information rates. For such inputs,
the entropy increase caused by the BSC can
be strictly positive, suggesting that
\eqref{eq:del_bsc_existing_bound} can become smaller.

To obtain a tighter bound, we revisit the
identity \eqref{eq:del_bsc_entropy_identity}.
Since $H(U\mid Y)$ is known exactly, obtaining
a tighter upper bound on $H(Y\mid U)$ reduces
to deriving a lower bound on $H(U)$ in terms
of $H(Y)$. This motivates the use of
Mrs.\ Gerber's Lemma \cite[Theorem 1]{wyner1973theorem}, which relates the entropy
of the output of a BSC to the entropy of its input.
In the following lemma, we apply Mrs.\ Gerber's
Lemma to the repetition-channel output,
conditioned on its length, to derive an
improved upper bound on $H(Y\mid U)$.
First, consider the following notation.

\begin{convention}
For $a,b\in[0,1]$, define
\begin{align}
a\star b
&:=a(1-b)+(1-a)b.
\end{align}
Moreover, let $h_2^{-1}:[0,1]\to[0,1/2]$
denote the inverse of the binary entropy
function restricted to $[0,1/2]$.
\end{convention}

\begin{lemma}
\label{lem:QPlusBSC}
Let $X^n$ be an arbitrary binary input sequence and let $Y$ be the output of a
$Q$-repetition channel, with
${0<\mu_Q:=\mathbb{E}[Q]<\infty}$. Let $U$ be obtained by passing each symbol of
$Y$ independently through a $\operatorname{BSC}(\epsilon)$, with
$0\leq\epsilon\leq \frac{1}{2}$. Let $M:=|Y|=|U|$ denote the output length and define
\begin{align}
\bar r_n:=\frac{H(Y\mid M)}{n\mu_Q}.
\label{eq:def_rbar}
\end{align}
Then, we have
\begin{align*}
\frac{1}{n}I(X^n;U)
\geq
\frac{1}{n}I(X^n;Y)
&-\mu_Q h_2(\epsilon)\\
&+\mu_Q\left[
h_2\!\left(\epsilon\star h_2^{-1}(\bar r_n)\right)-\bar r_n
\right].
\end{align*}
\end{lemma}

It can be shown (see Lemma~\ref{lem:gepsNonincreasing}
in the proof section) that the function
\begin{align*}
g_\epsilon(r)
:=h_2\!\left(\epsilon\star h_2^{-1}(r)\right)-r
\end{align*}
is nonnegative and nonincreasing on $[0,1]$.
Therefore, to obtain a lower bound on $g_\epsilon(r)$, and hence,
$\frac{1}{n}I(X^n;U)$, it suffices to derive an
asymptotic upper bound on $\frac{H(Y\mid M)}{n\mu_Q}$.

Since many existing capacity lower bounds for
repetition channels are obtained using first-order
Markov input processes, we focus on deriving such
an upper bound for this class of inputs.
The following lemma provides an explicit asymptotic
upper bound on the normalized conditional entropy
of the repetition-channel output.
\begin{table*}[t]
\centering
\caption{
Coefficients computed using the data from
Table~I of \cite{drinea2007improved}, crossover thresholds,
and capacity lower bounds with and without the
Mrs.~Gerber correction. The last column gives the
MGL lower bound evaluated at the baseline crossover
threshold.
}
\label{tab:DelBSC_thresholds}
\setlength{\tabcolsep}{3pt}
\begin{tabular}{|c|cc|cc|cc|cc|c|}
\hline
& & & 
\multicolumn{2}{c|}{Thresholds} &
\multicolumn{2}{c|}{$\epsilon=0.001$} &
\multicolumn{2}{c|}{$\epsilon=0.01$} &
$\epsilon=\epsilon_i^{\mathrm{base}}$ \\
\cline{4-10}
$d_i$ & $a_i$ & $b_i$
& $\epsilon_i^{\mathrm{MGL}}$
& $\epsilon_i^{\mathrm{base}}$
& $L_i^{\mathrm{MGL}}$
& $L_i^{\mathrm{base}}$
& $L_i^{\mathrm{MGL}}$
& $L_i^{\mathrm{base}}$
& $L_i^{\mathrm{MGL}}$ \\
\hline
0.05 & 0.471415 & 0.231021 & 0.224479 & 0.223563 & 0.717461 & 0.717452 & 0.651624 & 0.651536 & 0.001556 \\
0.10 & 0.436106 & 0.363789 & 0.158500 & 0.155896 & 0.551735 & 0.551692 & 0.489668 & 0.489246 & 0.005602 \\
0.15 & 0.394191 & 0.450769 & 0.120408 & 0.115635 & 0.429594 & 0.429483 & 0.371609 & 0.370505 & 0.011368 \\
0.20 & 0.354078 & 0.504295 & 0.095679 & 0.089071 & 0.337766 & 0.337564 & 0.284059 & 0.282055 & 0.016350 \\
0.25 & 0.314607 & 0.530581 & 0.078773 & 0.070514 & 0.267636 & 0.267323 & 0.218374 & 0.215284 & 0.020338 \\
0.30 & 0.274464 & 0.530041 & 0.067699 & 0.057579 & 0.214886 & 0.214444 & 0.170251 & 0.165874 & 0.023817 \\
0.35 & 0.242657 & 0.520955 & 0.059414 & 0.048148 & 0.174143 & 0.173594 & 0.133920 & 0.128494 & 0.024916 \\
0.40 & 0.219781 & 0.512418 & 0.052705 & 0.041108 & 0.142179 & 0.141565 & 0.106001 & 0.099934 & 0.023958 \\
0.45 & 0.194903 & 0.488235 & 0.048522 & 0.035945 & 0.117271 & 0.116585 & 0.085198 & 0.078423 & 0.023503 \\
0.50 & 0.180328 & 0.477075 & 0.044347 & 0.031877 & 0.096853 & 0.096156 & 0.068347 & 0.061463 & 0.021274 \\
0.55 & 0.163935 & 0.456257 & 0.041467 & 0.028612 & 0.079899 & 0.079189 & 0.054969 & 0.047966 & 0.019492 \\
0.60 & 0.161291 & 0.463480 & 0.037597 & 0.025999 & 0.065644 & 0.065001 & 0.043595 & 0.037246 & 0.016112 \\
0.65 & 0.158731 & 0.468814 & 0.034423 & 0.023833 & 0.053439 & 0.052865 & 0.034241 & 0.028580 & 0.013211 \\
0.70 & 0.156250 & 0.474183 & 0.031335 & 0.021736 & 0.042403 & 0.041901 & 0.026027 & 0.021086 & 0.010549 \\
0.75 & 0.129033 & 0.410845 & 0.032511 & 0.020446 & 0.033642 & 0.033132 & 0.020808 & 0.015786 & 0.010089 \\
0.80 & 0.120968 & 0.395807 & 0.031210 & 0.019095 & 0.025417 & 0.024984 & 0.015369 & 0.011107 & 0.008009 \\
0.85 & 0.108696 & 0.366771 & 0.030768 & 0.017851 & 0.018024 & 0.017669 & 0.010757 & 0.007261 & 0.006151 \\
0.90 & 0.118111 & 0.400127 & 0.027530 & 0.016920 & 0.011458 & 0.011237 & 0.006474 & 0.004299 & 0.003636 \\
0.95 & 0.110585 & 0.386858 & 0.025661 & 0.015409 & 0.005288 & 0.005171 & 0.002851 & 0.001701 & 0.001753 \\
\hline
\end{tabular}
\end{table*}




\begin{lemma}
\label{lem:HYMupperBound}
Let $X^n=(X_1,\ldots,X_n)$ be generated by a symmetric
first-order binary Markov chain with
\begin{align*}
\Pr(X_{i+1}=X_i)=\alpha,
\qquad 0\leq\alpha\leq1.
\end{align*}
Let $Y$ be the output of a $Q$-repetition channel
with $0<\mu_Q:=\mathbb{E}[Q]<\infty,$ and $q_0:=\Pr(Q=0)$.
Moreover, let $M:=|Y|$ denote the output length.
Then, we have
\begin{align*}
\limsup_{n\to\infty}
\frac{H(Y\mid M)}{n\mu_Q}
\leq h_2\left(\frac{(1-\alpha)(1-q_0)}
{\mu_Q[1-q_0(2\alpha-1)]}\right).
\end{align*}
\end{lemma}

We are now ready to combine
Lemmas~\ref{lem:QPlusBSC} and
\ref{lem:HYMupperBound} to obtain explicit
lower bounds on the capacity of the
$\Del(d)+\BSC(\epsilon)$ channel.
To this end, we use the achievable rates derived
in \cite{drinea2007improved,mitzenmacher2006simple},
which are based on symmetric first-order binary
Markov inputs. In particular, Mitzenmacher and
Drinea~\cite{mitzenmacher2006simple} established
a lower bound on the capacity of the Poisson
repetition channel using a Markov input with
$\alpha=0.77$ and Poisson parameter $\lambda=1.79$.

Furthermore, Table~I of
\cite{drinea2007improved} provides achievable
rates for deletion channels using first-order
Markov inputs with different values of $\alpha$.
By combining these achievable rates with
Lemmas~\ref{lem:QPlusBSC} and
\ref{lem:HYMupperBound}, we obtain lower bounds
on the capacity of the $\Del(d)+\BSC(\epsilon)$ channel.
We summarize these results in the following
corollary.

\begin{corollary}
\label{cor:DelBSCLowerBound}
For every $0\leq d<1$ and
$0\leq\epsilon\leq1/2$, the capacity of the
$\Del(d)+\BSC(\epsilon)$ channel satisfies
\begin{align}
C(&\Del(d)+\BSC(\epsilon))
\nonumber\\
&\quad\geq
(1-d)
\Big[
h_2(\epsilon\star 0.1177)
-h_2(\epsilon)
-0.4043
\Big].
\label{eq:DelBSCPoisLowerBound1}
\end{align}
Moreover, we have
\begin{align}
C(\Del(d_i)+\BSC(\epsilon))
\geq
(1-d_i)
\Big[
h_2(\epsilon\star a_i)
-h_2(\epsilon)
-b_i
\Big],
\label{eq:DelBSCPoisLowerBound2}
\end{align}
where the coefficients $a_i$ and $b_i$ are
given in Table~\ref{tab:DelBSC_thresholds}.
\end{corollary}

Corollary~\ref{cor:DelBSCLowerBound} provides two types
of lower bounds on the capacity of the
$\Del(d)+\BSC(\epsilon)$ channel.
The first bound, given in
\eqref{eq:DelBSCPoisLowerBound1}, is valid for
all $d\in[0,1)$ and is particularly useful
in the high-deletion regime.
The second bound, given in
\eqref{eq:DelBSCPoisLowerBound2}, provides
improved lower bounds at the specific deletion
probabilities $d_i$ listed in
Table~\ref{tab:DelBSC_thresholds}.
However, these numerical bounds can also be
extended to the entire interval $d\in[0,0.95]$.
As established in \cite{rahmati2014upper},
the normalized capacity
\begin{align*}
\frac{C(\Del(d)+\BSC(\epsilon))}{1-d}
\end{align*}
is nonincreasing in $d$.
This property can also be derived using the
repetition-based approach, following an argument
similar to that used in
Corollary~\ref{cor:CDelnonincreasing}.

Furthermore, since
$h_2\!\left(\epsilon\star h_2^{-1}(r)\right)-r$
is nonnegative for $0\leq\epsilon\leq1/2$,
our approach provides a lower bound that is never weaker
than the corresponding bound in
\cite{rahmati2014achievable}.
To quantify this improvement,
Table~\ref{tab:DelBSC_thresholds} presents numerical
comparisons based on the coefficients $a_i$ and $b_i$
computed using the data from Table~I of
\cite{drinea2007improved}.
For each deletion probability $d_i$, we define
$\epsilon_i^{\mathrm{MGL}}$ and
$\epsilon_i^{\mathrm{base}}$ as the largest BSC crossover
probabilities for which the lower bounds with and
without the Mrs.\ Gerber correction, respectively,
remain positive. 

The table also reports the capacity lower bounds
$L_i^{\mathrm{MGL}}$ and $L_i^{\mathrm{base}}$
obtained with and without the Mrs.\ Gerber correction,
respectively, for the fixed crossover probabilities
$\epsilon=0.001$ and $\epsilon=0.01$.
The results show that the Mrs.\ Gerber correction
improves the lower bound for every deletion probability
considered, with a more pronounced relative improvement
at larger crossover probabilities.
The last column reports
$L_i^{\mathrm{MGL}}(\epsilon_i^{\mathrm{base}})$,
namely, the lower bound with the Mrs.\ Gerber correction
evaluated at the crossover threshold of the baseline bound.
At this crossover probability, the baseline lower bound
vanishes, whereas the corrected bound remains strictly
positive for every deletion probability listed in the table.

The improvement is particularly pronounced in the
high-deletion regime. For example, at $d=0.95$,
the crossover threshold increases from approximately
$0.0154$ without the correction to $0.0257$ with
the correction. This behavior is related to the
statistical structure of the Markov inputs used to
obtain the achievable rates in
\cite{drinea2007improved}.
At higher deletion probabilities, the input distributions
considered in \cite{drinea2007improved} generally have
larger values of $\alpha=\Pr(X_{i+1}=X_i)$,
corresponding to longer input runs and stronger
correlations between consecutive bits. These
correlations allow Mrs.\ Gerber's Lemma to provide
a larger entropy correction, leading to more
substantial improvements over the baseline lower
bounds. In contrast, when $\alpha=1/2$, the input
bits are i.i.d.\ and uniformly distributed, and
the output of the deletion channel, conditioned on
its length, is also i.i.d.\ and uniformly distributed.
Consequently, the entropy correction vanishes,
and Mrs.\ Gerber's Lemma provides no improvement
over the baseline bound in this case.

\section{Proofs of the Deletion Channel Results}\label{sec:proof_del}
In this section, we provide the proofs of the results presented in Section \ref{sec:main_results}.
\subsection{Proofs for Repetition before the Channel}

\begin{proof}[Proof of Lemma \ref{lem:MainPostEncoding}]
To prove the lemma, we follow an approach similar to that
of~\cite{mitzenmacher2006simple}. Fix any $R<C(Q')$. By the definition
of an achievable rate, there exists a sequence of length-$k$ codes for
the $Q'$-repetition channel with message set $\mathcal W_k=\{1,\ldots,2^{\lfloor kR\rfloor}\},$
a fixed-length encoder
$\psi_k:\mathcal W_k\to\{0,1\}^k$, and a decoder $\phi_k$, such that
the average decoding error probability $\epsilon_k\to0$ as
$k\to\infty$.

We construct from this code a code for the $Q$-repetition channel. For
every bit of the codeword
$\psi_k(w)=x_1(w)\cdots x_k(w)$ corresponding to message
$w\in\mathcal W_k$, independently generate $A_i\sim P$,
$i=1,\ldots,k$, and replace $x_i(w)$ by $A_i$ consecutive copies.
The resulting random codeword is
\begin{align}\label{eqn:PsiEncoder}
\Psi_k(w)
=
x_1(w)^{A_1}\cdots x_k(w)^{A_k},
\end{align}
and has random length $V=\sum_{i=1}^k A_i.$
Since the $A_i$'s are i.i.d., we have 
\begin{align}\label{eqn:EVarV}
\mathbb E[V]
=
k\mathbb{E}[A], \quad\Var(V)
=
k\Var(A).
\end{align}

We first observe that the average error probability of this random code
over the $Q$-repetition channel is exactly $\epsilon_k$. Consider one original bit $x_i$. After the first repetition stage, it
is replaced by $A_i$ copies. Let
$B_{i,1},\ldots,B_{i,A_i}$ denote the independent numbers of copies
produced by the $Q$-repetition channel from these intermediate copies.
The total number of copies of $x_i$ appearing at the final output is
therefore $D_i=\sum_{j=1}^{A_i} B_{i,j}.$
By assumption, we have $D_i\sim Q'.$
Moreover, $D_1,\ldots,D_k$ are independent because all the repetitions are
performed independently across the original input bits. Therefore, the
cascade of the $P$-repetition stage and the $Q$-repetition channel is
exactly the $Q'$-repetition channel. Applying the decoder $\phi_k$ to
the output of the cascade consequently gives average decoding error
probability $\epsilon_k\to0$. Hence, $\Psi_k$, defined in \eqref{eqn:PsiEncoder}, together with $\phi_k$
defines a random-length code for the $Q$-repetition channel with
average error probability $\epsilon_k$.

It remains to replace the random-length codewords by codewords having
a common fixed length. 
Define
\begin{align*}
\mathcal T_k
:=
\left\{
v\in\mathbb Z_{\geq0}:
|v-k\mathbb{E}[A]|\leq k^{3/4}
\right\}.
\end{align*}
By Chebyshev's inequality, we have
\begin{align*}
\Pr(V\notin\mathcal T_k)
&\leq
\frac{\Var(V)}{k^{3/2}}\\
&=
\frac{\Var(A)}{k^{1/2}},
\end{align*}
which follows from \eqref{eqn:EVarV}.
Therefore, we have $\Pr(V\in\mathcal T_k)\to 1$ as $k\to \infty$.

Let $\epsilon_k(v)$ denote the average decoding error probability of
the randomized code $\Psi_k$ conditioned on its codeword having total
length $v$. The law of total probability implies 
\begin{align*}
\epsilon_k
&=
\sum_v \Pr(V=v)\epsilon_k(v)\\
&\geq
\sum_{v\in\mathcal T_k}
\Pr(V=v)\epsilon_k(v).
\end{align*}
Therefore, there exists some $v_k^*\in\mathcal T_k$ with
$\Pr(V=v_k^*)>0$ such that
\begin{align*}
\epsilon_k(v_k^*)
\leq
\frac{\epsilon_k}
{\Pr(V\in\mathcal T_k)}.
\end{align*}
Since $\epsilon_k\to0$ and
$\Pr(V\in\mathcal T_k)\to1$, we obtain $\epsilon_k(v_k^*)\to 0.$
Moreover, because $v_k^*\in\mathcal T_k$, we have 
\begin{align*}
\left|
\frac{v_k^*}{k}-\mathbb{E}[A]
\right|
&\leq
k^{-1/4},
\end{align*}
and therefore $\frac{v_k^*}{k}
\to
\mathbb{E}[A].$

We now generate the expansion of every message conditioned on having
total length $v_k^*$. The resulting randomized encoder has the same
message set $\mathcal W_k$, fixed blocklength $v_k^*$, and average
decoding error probability $\epsilon_k(v_k^*)$. Averaging over the
randomness of the encoder, there must exist a deterministic choice of
the expanded codewords whose average decoding error probability is no
greater than $\epsilon_k(v_k^*)$. Hence, there exists a deterministic
fixed-length code for the $Q$-repetition channel with
$2^{\lfloor kR\rfloor}$ messages, blocklength $v_k^*$, and average
error probability tending to zero.

The rate of this code satisfies
\begin{align*}
\frac{\lfloor kR\rfloor}{v_k^*}
=
\frac{\lfloor kR\rfloor/k}{v_k^*/k}
\to
\frac{R}{\mathbb{E}[A]}.
\end{align*}
It follows that $C(Q)
\geq
\frac{R}{\mathbb{E}[A]}.$
Since $R<C(Q')$ was arbitrary, we conclude that
\begin{align*}
C(Q)
\geq
\frac{1}{\mathbb{E}[A]}C(Q'),
\end{align*}
which completes the proof.
\end{proof}


\begin{proof}[Proof of Corollary \ref{thm:bin_lower_del}]
Let $P=\Bin\!\left(m,p/(1-d)\right)$ and
${Q=\Ber(1-d)}$. Since $d\leq1-p$, we have
$p/(1-d)\leq1$, so $P$ is well defined.

Let $A\sim P$ and, independently, let
$B_1,B_2,\ldots\sim\Ber(1-d)$ be i.i.d. We can write
$A=\sum_{i=1}^m U_i$, where
$U_i\sim\Ber\!\left(\frac{p}{1-d}\right)$ are i.i.d. Then
\begin{align*}
\sum_{j=1}^{A}B_j
\sim
\sum_{i=1}^m U_iV_i,
\end{align*}
where $V_1,\ldots,V_m$ are i.i.d.\ $\Ber(1-d)$ random variables
independent of the $U_i$'s. Since
$U_iV_i\sim\Ber(p)$ independently, we obtain
$\sum_{j=1}^{A}B_j\sim Q'=\Bin(m,p)$.
Also, $\mathbb E[A]=mp/(1-d)$. Therefore,
Lemma~\ref{lem:MainPostEncoding} completes the proof.
\end{proof}

\begin{proof}[Proof of Corollary \ref{cor:CDelnonincreasing}]
From Corollary \ref{thm:bin_lower_del}, we have
\begin{align*}
\frac{C(\Del(d))}{1-d}
\ge
\frac{C(\Bin(m,p))}{mp},
\qquad d\le1-p.
\end{align*}
 Moreover, $\BinCh(1,p)$-repetition channel is exactly a deletion channel with deletion probability $d'=1-p$. Hence, we have
\begin{align*}
\frac{C(\Del(d))}{1-d}
\ge
\frac{C(\Del(d'))}{1-d'}.
\end{align*}
Finally, $d'=1-p\geq d$ completes the proof.
\end{proof}
\subsection{Proofs of Repetition after the Channel (Data Processing Inequality)}

To prove Theorem~\ref{thm:bin_pois_comparison}, we use an additional
repetition stage after the binomial-repetition channel, with the
following repetition distribution.
\begin{definition}
Suppose that
\begin{equation}
    e^{-\lambda/m}-(1-p)\geq 0.
    \label{eq:condition}
\end{equation}
    Define non-negative integer Binomial to Poisson distribution $\Btp(m,p,\lambda)$ as follow
    \begin{align}
    \Pr(R=k)
    =\begin{cases}
    \frac{e^{-\lambda/m}-(1-p)}{p},  &\text{if }k=0\\ \\
       \frac{e^{-\lambda/m}}{p}
    \frac{(\lambda/m)^k}{k!}
    ,  &\text{if }k>0
    \end{cases}
    \label{eq:BtpPmf}
    \end{align}
\end{definition}

The nonnegativity of the probabilities follows
from the fact that $p\geq1-e^{-\lambda/m}$. Moreover, we have
\begin{align*}
    \sum_{r=0}^{\infty}\Pr(R=r)
    &=-\frac{1-p}{p}+\frac{1}{p}\sum_{r=0}^{\infty}\frac{(\lambda/m)^re^{-\lambda/m}}{r!} \nonumber\\
    &\stackrel{(a)}{=}-\frac{1-p}{p}+\frac{1}{p} 
    =1,
\end{align*}
where $(a)$ follows from the fact that $\frac{(\lambda/m)^re^{-\lambda/m}}{r!}$ is the probability mass function (pmf) of a Poisson distribution with mean $\frac{\lambda}{m}$.
Thus, $\Btp(m,p,\lambda)$ defines a valid
probability distribution.


To characterize the distribution of the cascade of the binomial
repetition channel and the additional repetition stage, we use
probability generating functions (PGFs). For a nonnegative
integer-valued random variable $Z$, its PGF is defined by
\begin{align*}
G_Z(z)
=
\mathbb E[z^Z]
=
\sum_{k=0}^{\infty}\Pr(Z=k)z^k,
\end{align*}
for values of $z$ for which the expectation exists
\cite[Ch.~XI]{feller1991introduction}. More specifically, we use the
following standard composition property.

\begin{theorem}[PGF of a Random Sum {\cite[Ch.~XII.1]{feller1991introduction}}]
\label{lem:pgf_comp}
Let $N$ be a nonnegative integer-valued random variable, and let
$R_1,R_2,\ldots$ be i.i.d.\ nonnegative integer-valued random
variables independent of $N$. Define
\begin{align*}
S
=
\sum_{j=1}^{N} R_j,
\end{align*}
where the sum is taken to be zero when $N=0$. Then
\begin{align}\label{eq:GY}
G_S(z)
=
G_N\!\left(G_R(z)\right),
\end{align}
where $G_R$, $G_N$, and $G_S$ are the PGF of each $R$, $N$, and $S$, respectively.
\end{theorem}

We are now ready to characterize the distribution of the cascade of a
$\Bin(m,p)$ repetition channel with a $\Btp(m,p,\lambda)$ repetition
channel, as established in the following lemma.

\begin{lemma}\label{lem:BinToPois}
Let $K\sim\Bin(m,p)$ and independently $R_i\sim\operatorname{Btp}(m,p,\lambda)$ for $i=1,\ldots,K$. Suppose that
\begin{equation*}
    e^{-\lambda/m}-(1-p)\geq 0.
\end{equation*}
Then $Y=\sum_{i=1}^{K}R_i$ has a $\Pois(\lambda)$ distribution.
\end{lemma}

\begin{proof}

The PGF of $R$ is
\begin{align}
    G_R(z)
    &=\frac{e^{-\lambda/m}-(1-p)}{p}
    +\frac{e^{-\lambda/m}}{p}\sum_{r=1}^{\infty}\frac{(\lambda z/m)^r}{r!}\nonumber\\
    &\stackrel{(a)}{=}\frac{e^{-\lambda/m}-(1-p)}{p}
    +\frac{e^{-\lambda/m}(e^{\lambda z/m}-1)}{p} \nonumber\\
    &=\frac{\exp\left(\frac{\lambda}{m}(z-1)\right)-(1-p)}{p}.\label{eq:GR}
\end{align}
where $(a)$ follows from the Taylor series of $\exp(x)-1=\sum_{r=1}^{\infty}\frac{x^r}{r!}$.
Moreover, the PGF of a binomial random variable is \cite[Ch.~XI.2]{feller1991introduction}
\begin{align}
    G_K(u)=(1-p+pu)^m.
    \label{eq:GK}
\end{align}
To compute the PGF of the total number of copies $Y$, we apply
Theorem~\ref{lem:pgf_comp}. Substituting \eqref{eq:GR} and
\eqref{eq:GK} into \eqref{eq:GY}, we obtain
\begin{align}
    G_Y(z)
    &=\left(1-p+pG_R(z)\right)^m \nonumber\\
    &=\left[1-p+p\frac{\exp\left(\frac{\lambda}{m}(z-1)\right)-(1-p)}{p}\right]^m \nonumber\\
    &=\left[\exp\left(\frac{\lambda}{m}(z-1)\right)\right]^m \nonumber\\
    &=\exp\left(\lambda(z-1)\right).
    \label{eq:GY_final}
\end{align}

The PGF of a Poisson random variable with parameter $\lambda$ is $G_{\operatorname{Pois}(\lambda)}(z)=\exp\left(\lambda(z-1)\right)$ \cite[Ch.~XI.2]{feller1991introduction}.
Comparing this expression with \eqref{eq:GY_final}, we conclude that $Y$ and a $\operatorname{Pois}(\lambda)$ random variable have the same PGF. Since the PGF uniquely determines the probability mass function of a nonnegative integer-valued random variable \cite[Ch.~XI.4]{feller1991introduction}, it follows that $Y\sim\Pois(\lambda).$
\end{proof}

\begin{proof}[Remark]
As shown in Lemma~\ref{lem:BinToPois}, a binomial repetition channel
can be followed by a suitable repetition channel to produce exactly a
Poisson repetition channel. One might ask whether an analogous
construction is possible in the reverse order, by placing an additional
repetition stage before the binomial channel, as in the previous
subsection. In general, however, such a construction is not possible.

To see this, suppose that the secondary repeat channel is binomial rather than the first one. More precisely, let $X_1,X_2,\ldots\overset{\mathrm{i.i.d.}}{\sim}\operatorname{Bin}(m,p)$, and let $K$ be an arbitrary nonnegative integer-valued random variable independent of the $X_i$'s. Consider the total repetition count $Y=\sum_{i=1}^{K}X_i$. We ask whether it is possible to choose the distribution of $K$ such that $Y\sim\Pois(\lambda)$.

The PGF of each $X_i$ is $G_X(z)=(1-p+pz)^m$. Since $Y$ is a random sum, Theorem \ref{lem:pgf_comp} implies
\begin{equation*}
    G_Y(z)=G_K(G_X(z))=G_K\left((1-p+pz)^m\right).
\end{equation*}
If $Y\sim\Pois(\lambda)$, then
\begin{equation*}
    G_K\left((1-p+pz)^m\right)=\exp\left(\lambda(z-1)\right).
\end{equation*}
Letting $u=(1-p+pz)^m$, we have $z=[u^{1/m}-(1-p)]/p$, and consequently $z-1=(u^{1/m}-1)/p$. Therefore, the only possible candidate for the PGF of $K$ is
\begin{equation}
    G_K(u)=\exp\left(\frac{\lambda}{p}\left(u^{1/m}-1\right)\right).
    \label{eq:reverse_candidate}
\end{equation}

We next show that \eqref{eq:reverse_candidate} cannot be the PGF of a nonnegative integer-valued random variable when $m>1$.  Taking derivative of $G_K(u)$, we have
\begin{equation}
    G_K'(u)=\frac{\lambda}{mp}u^{1/m-1}\exp\left(\frac{\lambda}{p}\left(u^{1/m}-1\right)\right).
\end{equation}
For $m>1$, we have  $u^{1/m-1}\to\infty$ as $u\to0$. At the same time, $\exp(\frac{\lambda}{p}(u^{1/m}-1))\rightarrow e^{-\frac{\lambda}{p}}>0$. It follows that
\begin{equation}\label{eqn:GKinfty}
    G_K'(0)=+\infty.
\end{equation}

On the other hand, if $K$ is any nonnegative integer-valued random variable, then 
\begin{align*}
G_K'(0)
=
\left.
\sum_{k=1}^{\infty}
k\,\Pr(K=k)z^{k-1}
\right|_{z=0}
=
\Pr(K=1),
\end{align*}
which must be finite and belong to $[0,1]$. This contradicts \eqref{eqn:GKinfty}. Therefore, \eqref{eq:reverse_candidate} cannot be a valid PGF for $m>1$, and no nonnegative integer-valued distribution for $K$ can make $\sum_{i=1}^{K}X_i$ exactly Poisson when $X_i\overset{\mathrm{i.i.d.}}{\sim}\operatorname{Bin}(m,p)$ with $m>1$.

For $m=1$, however, such a construction is possible. In this case,
$\Bin(1,p)$ is a Bernoulli distribution, and a suitable Poisson
repetition stage before the Bernoulli repetition channel produces a
Poisson repetition channel \cite{mitzenmacher2006simple}.
\end{proof}

\begin{proof}[Proof of Theorem \ref{thm:bin_pois_comparison}]
By Lemma~\ref{lem:BinToPois}, under the condition
$e^{-\lambda/m}-(1-p)\geq0$, a
$\operatorname{Pois}(\lambda)$ repetition channel can be obtained by
cascading a $\operatorname{Bin}(m,p)$ repetition channel with a
$\operatorname{Btp}(m,p,\lambda)$ repetition channel. Hence,
$X^n\longrightarrow Y\longrightarrow Z$ forms a Markov chain, where
$Y$ and $Z$ are the outputs of the binomial and Poisson repetition
channels, respectively. 

For every input distribution $P_{X^n}$,
the Markov chain $X^n\to Y\to Z$ and
the data-processing inequality imply
\begin{align*}
I(X^n;Z)\leq I(X^n;Y).
\end{align*}
Let $P_{X^{n,*}}$ be an input distribution
that maximizes $I(X^n;Z)$, and let
$X^{n,*}$ denote an input sequence
generated according to $P_{X^n}^\star$.
Then, we have
\begin{align*}
I(X^{n,*};Z)\leq I(X^{n,*};Y)
\leq \max_{P_{X^n}} I(X^n;Y).
\end{align*}
Dividing by $n$ and letting $n\to\infty$
yields
\begin{align*}
C(\Pois(\lambda))&=\lim_{n\to \infty}\frac{1}{n}I(X^{n,*};Z)\cr
&\leq \lim_{n\to \infty}\frac{1}{n}\max_{P_{X^n}} I(X^n;Y)=C(\Bin(m,p)).
\end{align*}
\end{proof}

\begin{proof}[Proof of Corollary \ref{cor:UpperBoundPoissonCh}]
To combine $C(\Del(d))\le \alpha(1-d)$ for all
$d\ge \beta$ with Corollary~\ref{cor:del_lower_bound}, we need a $d$
satisfying
$\beta\leq d\leq e^{-\lambda/m}.$
Such a $d$ exists  when
$e^{-\lambda/m}\geq\beta,$
or equivalently when
$m\geq \frac{\lambda}{-\log(\beta)}.$
For every such $m$, choosing $d=\beta$ gives
\begin{align*}
\frac{C(\Pois(\lambda))}
{m(1-e^{-\lambda/m})}
\leq
\frac{C(\Del(\beta))}{1-\beta}
\leq \alpha.
\end{align*}
Hence
$C(\Pois(\lambda))
\leq
\alpha\,m(1-e^{-\lambda/m}).$

Finally, since $m(1-e^{-\lambda/m})$ is increasing in $m$, the
strongest bound is obtained by the smallest admissible positive integer,
which proves the result.
\end{proof}

\subsection{Proofs of Repetition after the Channel (Beyond Data Processing Inequality)}
\begin{proof}[Proof of Lemma \ref{lem:LimitPoissonOverLambda}]
First, note that $C(\Pois(\lambda))$ is nondecreasing in $\lambda$.
Indeed, if $0<\lambda_1\leq\lambda_2$, then cascading a
$\Pois(\lambda_2)$ repetition channel with a
$\Ber(\lambda_1/\lambda_2)$ repetition channel produces a
$\Pois(\lambda_1)$ repetition channel. Hence, by the data-processing
inequality, we have
\begin{align}\label{eqn:montonocityCPois}
C(\Pois(\lambda_1))
\leq
C(\Pois(\lambda_2)).
\end{align}

From the definitions of the supremum and the $\limsup$, we have
\begin{align}\label{eqn:limsuplessSup}
\limsup_{\mu\to0}
\frac{C(\Pois(\mu))}{\mu}
\leq
\sup_{\lambda>0}
\frac{C(\Pois(\lambda))}{\lambda}.
\end{align}

To prove the reverse inequality for the $\liminf$, fix any
$\lambda>0$ and let
$m_\mu=\lceil\lambda/\mu\rceil$. Then
$\lambda\leq m_\mu\mu<\lambda+\mu$. From \eqref{eqn:montonocityCPois}, we have
\begin{align*}
C(\Pois(\lambda))
&\leq
C(\Pois(m_\mu\mu))\cr
&\leq
m_\mu C(\Pois(\mu)),
\end{align*}
where the last inequality comes form Corollary~\ref{cor:PoisChScaling}.
Therefore, we have
\begin{align*}
\frac{C(\Pois(\mu))}{\mu}
\geq
\frac{\lambda}{m_\mu\mu}
\frac{C(\Pois(\lambda))}{\lambda}.
\end{align*}
Since $\lambda/(m_\mu\mu)\to1$ as $\mu\to0$, we obtain
\begin{align*}
\liminf_{\mu\to0}
\frac{C(\Pois(\mu))}{\mu}
\geq
\frac{C(\Pois(\lambda))}{\lambda}.
\end{align*}
Since $\lambda>0$ is arbitrary, we have
\begin{align*}
\liminf_{\mu\to0}
\frac{C(\Pois(\mu))}{\mu}
\geq
\sup_{\lambda>0}
\frac{C(\Pois(\lambda))}{\lambda}.
\end{align*}
Combining this  with \eqref{eqn:limsuplessSup}
implies the existence of the limit
of $C(\Pois(\lambda))/\lambda$ as $\lambda\to0$
and yields
\begin{align*}
\lim_{\lambda\to0}
\frac{C(\Pois(\lambda))}{\lambda}
=
\sup_{\lambda>0}
\frac{C(\Pois(\lambda))}{\lambda},
\end{align*}
which completes the proof.

\end{proof}

\begin{lemma}\label{lem:BtpEntropy}
For every fixed integer $m\geq1$, we have
\begin{align*}
\lim_{\lambda\to0}
H\!\left(
\Btp\!\left(m,1-e^{-\lambda/m},\lambda\right)
\right)
=0,
\end{align*}
where $H(\cdot)$ denotes the  entropy.
\end{lemma}
\begin{proof}
Let
\begin{align*}
R_\lambda
\sim
\Btp\!\left(m,1-e^{-\lambda/m},\lambda\right),
\end{align*}
and define $q_\lambda(k):=\Pr(R_\lambda=k)$. 
Then $q_\lambda(0)=0$, and for every $k\geq1$, we have
\begin{align}\label{eqn:qlambdak}
q_\lambda(k)
=
\frac{(\lambda/m)^k}{k!(e^{\lambda/m}-1)}.
\end{align}
By the definition of entropy, we have
\begin{align}\label{eqn:EnropyBtp}
H(R_\lambda)
=
\sum_{k=1}^{\infty}
q_\lambda(k)\log_2\frac{1}{q_\lambda(k)}.
\end{align}

For every fixed $k\geq1$, as $\lambda\to0$, 
$q_\lambda(1)\to1$ and $q_\lambda(k)\to0$ for $k\geq2$. Hence, for all $k$, we have
\begin{align*}
q_\lambda(k)\log_2\frac{1}{q_\lambda(k)}
\to 0.
\end{align*}
Therefore, it remains to justify interchanging the limit and the sum in \eqref{eqn:EnropyBtp}. Let $x=\frac{\lambda}{m}.$
 Since $\lambda\to 0$, restrict
$0<x\leq 1$. Since for all $x\geq 0$, $e^x-1\geq x$, \eqref{eqn:qlambdak} gives
\begin{align}\label{eqn:qUpper}
q_\lambda(k)
\leq
\frac{x^{k-1}}{k!}.
\end{align}
Moreover, by the convexity of the exponential function, we have
$e^x-1\leq(e-1)x$ for $0<x\leq1$. Hence, \eqref{eqn:qlambdak} gives
\begin{align}\label{eqn:qLower}
q_\lambda(k)
\geq
\frac{x^{k-1}}{(e-1)k!}.
\end{align}
Combining \eqref{eqn:qUpper} and \eqref{eqn:qLower} gives
\begin{align*}
q_\lambda(k)\log_2\frac{1}{q_\lambda(k)}
&\leq
\frac{x^{k-1}}{k!}
\log_2\left(
\frac{(e-1)k!}{x^{k-1}}
\right)\\
&=
\frac{x^{k-1}}{k!}
\log_2\!\bigl((e-1)k!\bigr)
+
\frac{x^{k-1}}{k!}
\log_2\frac{1}{x^{k-1}}.
\end{align*}
Since $x^{k-1}\leq1$ and $x^{k-1}\log_2\frac{1}{x^{k-1}}
\leq 1$,
we obtain
\begin{align*}
q_\lambda(k)\log_2\frac{1}{q_\lambda(k)}
\leq
\frac{1}{k!}
\left(
\log_2\!\bigl((e-1)k!\bigr)
+
1
\right):=a_k.
\end{align*}
The sequence $a_k$
is summable by the ratio test, since  $\lim_{k\to \infty}\frac{a_{k+1}}{a_k}=0$.
 Therefore, by the dominated convergence theorem \cite[Theorem~2.24]{folland1999real}, we may interchange
the limit and the infinite sum, and hence
\begin{align*}
\lim_{\lambda\to0}
H\!&\left(
\Btp\!\left(m, 1-e^{-\lambda/m},\lambda\right)\right)\cr
&=
\sum_{k=1}^{\infty}
\lim_{\lambda\to0}
q_\lambda(k)\log_2\frac{1}{q_\lambda(k)}=0,
\end{align*}
which completes the proof.
\end{proof}

\begin{proof}[Proof of Theorem \ref{thm:BinPoisAsymptotic}]
For $\lambda>0$, let
$p_\lambda=1-e^{-\lambda/m}$. From 
Theorem~\ref{thm:bin_pois_comparison}, we know that $C(\Bin(m,p_\lambda))
\geq C(\Pois(\lambda)).$
Therefore, we have
\begin{align}\label{eqn:CBinmpCPoisL1}
\frac{C(\Bin(m,p_\lambda))}{mp_\lambda}
\geq
\frac{\lambda}{mp_\lambda}
\frac{C(\Pois(\lambda))}{\lambda}.
\end{align}
Since
\begin{align*}
\lim_{\lambda\to 0}\frac{mp_\lambda}{\lambda}
=
\lim_{\lambda\to 0}\frac{m(1-e^{-\lambda/m})}{\lambda}
= 1
\end{align*}
Taking the $\liminf$ in \eqref{eqn:CBinmpCPoisL1} and using
Lemma~\ref{lem:LimitPoissonOverLambda}, we obtain
\begin{align}\label{eqn:firstsideBinPois}
\liminf_{\lambda\to0}
\frac{C(\Bin(m,p_\lambda))}{mp_\lambda}
\geq
\lim_{\lambda\to0}
\frac{C(\Pois(\lambda))}{\lambda}.
\end{align}

We next prove the reverse inequality. Consider a
$\Bin(m,p_\lambda)$ repetition channel followed by a
$\Btp(m,p_\lambda,\lambda)$ repetition channel. By Lemma \ref{lem:BinToPois}, the
resulting channel is a $\Pois(\lambda)$ repetition channel. Let $X^n$
be an arbitrary input, let $Y$ denote the output of the Binomial-repetition 
channel, and let $Z$ denote the final Poisson-repetition channel output. Then
$X^n\to Y\to Z$ forms a Markov chain.
Therefore, we have $I(X^n;Z\mid Y)=0$, and consequently
\begin{align*}
I(X^n;Y,Z)
&=I(X^n;Y)+I(X^n;Z\mid Y)\\
&=I(X^n;Y).
\end{align*}
Using the chain rule again, we have
\begin{align*}
I(X^n;Y,Z)
&=I(X^n;Z)+I(X^n;Y\mid Z),
\end{align*}
Therefore, 
\begin{align*}
I(X^n;Y)
&=I(X^n;Z)+I(X^n;Y\mid Z),
\end{align*}
which gives
\begin{align}\label{eqn:IXYHYZ}
I(X^n;Y)-I(X^n;Z)=I(X^n;Y\mid Z)\leq H(Y\mid Z).
\end{align}
Let $U=(M,R_1,\ldots,R_M),$
where $M=|Y|$ and $R_1,\ldots,R_M$ are the repetition counts used by
the $\Btp(m,p_\lambda,\lambda)$ channel. Since
$p_\lambda=1-e^{-\lambda/m}$, we have
$\Pr(R_i=0)=0$. Hence, given $Z$ and $U$, the string $Y$ can be
reconstructed by partitioning $Z$ into consecutive blocks of lengths
$R_1,\ldots,R_M$. Therefore, we have 
\begin{align}\label{eqn:HYmidZU}
H(Y\mid Z,U)=0.
\end{align}
Using the chain rule, 
\begin{align}\label{eqn:HYZ}
H(Y\mid Z)
&\leq
H(Y,U\mid Z)\cr
&=
H(U\mid Z)+H(Y\mid Z,U)\cr
&\stackrel{(a)}{=}
H(U\mid Z)\cr
&\leq
H(U)\cr
&=
H(M)+H(R_1,\ldots,R_M\mid M),
\end{align}
where $(a)$ follows from \eqref{eqn:HYmidZU}.
Conditioned on $M=r$, the random variables $R_1,\ldots,R_r$ are
i.i.d.\ with distribution $\Btp(m,p_\lambda,\lambda)$. Hence,
\begin{align*}
H(R_1,&\ldots,R_M\mid M)\\
&=
\sum_{r\geq0}\Pr(M=r)\,
H(R_1,\ldots,R_r\mid M=r)\\
&=
\sum_{r\geq0}\Pr(M=r)\,
r\,H\!\left(\Btp(m,p_\lambda,\lambda)\right)\\
&=
\mathbb{E}[M]\,
H\!\left(\Btp(m,p_\lambda,\lambda)\right).
\end{align*}
Plugging this into \eqref{eqn:HYZ} gives
\begin{align}\label{eqn:HYZ2}
H(Y\mid Z)
\leq
H(M)
+
\mathbb{E}[M]\,
H\!\left(\Btp(m,p_\lambda,\lambda)\right).
\end{align}
Since the output length of the binomial channel satisfies
$M\sim\Bin(nm,p_\lambda)$, we have 
\begin{align*}
\mathbb{E}[M]
&=
nmp_\lambda.
\end{align*}
Combining this, \eqref{eqn:IXYHYZ}, and \eqref{eqn:HYZ2}, for every input distribution, we can write
\begin{align*}
I(X^n;Y)
&\leq
I(X^n;Z)+
H(M)
+
nmp_\lambda
H\!\left(\Btp(m,p_\lambda,\lambda)\right).
\end{align*}
Let $P_{X^{n,*}}$ be an input distribution that maximizes
$I(X^n;Y)$. Then
\begin{align*}
\max_{P_{X^n}}& I(X^n;Y)\\
&=
I(X^{n,*};Y)\\
&\leq
I(X^{n,*};Z)
+
H(M)
+
nmp_\lambda
H\!\left(\Btp(m,p_\lambda,\lambda)\right)\\
&\leq
\max_{P_{X^n}} I(X^n;Z)
+
H(M)
+
nmp_\lambda
H\!\left(\Btp(m,p_\lambda,\lambda)\right).
\end{align*}
Dividing by $n$ and letting $n\to \infty$ give
\begin{align*}
C(\Bin(m,p_\lambda))
\leq
C(\Pois(\lambda))
+
mp_\lambda
H\!\left(\Btp(m,p_\lambda,\lambda)\right),
\end{align*}
which follows since $M\in\{0,\ldots,nm\}$, and hence 
\begin{align*}
\lim_{n\to\infty}\frac{H(M)}{n}\leq \lim_{n\to\infty}\frac{\log_2(nm+1)}{n}=0,
\end{align*}
Therefore, we have
\begin{align*}
\frac{C(\Bin(m,p_\lambda))}{mp_\lambda}
\leq
\frac{\lambda}{mp_\lambda}
\frac{C(\Pois(\lambda))}{\lambda}
+
H\!\left(\Btp(m,p_\lambda,\lambda)\right).
\end{align*}
By Lemma~\ref{lem:BtpEntropy},
and $\lambda/(mp_\lambda)\to1$, we have
\begin{align*}
\limsup_{\lambda\to0}
\frac{C(\Bin(m,p_\lambda))}{mp_\lambda}
\leq
\lim_{\lambda\to0}
\frac{C(\Pois(\lambda))}{\lambda}.
\end{align*}
Combining this inequality and  \eqref{eqn:firstsideBinPois}  yields
\begin{align*}
\lim_{\lambda\to0}
\frac{C(\Bin(m,p_\lambda))}{mp_\lambda}
=
\lim_{\lambda\to0}
\frac{C(\Pois(\lambda))}{\lambda}.
\end{align*}

Finally, the mapping $p_\lambda=1-e^{-\lambda/m}$
is a bijection from $(0,\infty)$ onto $(0,1)$, with inverse $\lambda=-m\log(1-p).$
Moreover, $p\to0$ if and only if $\lambda\to0$. Therefore,
\begin{align*}
\lim_{\lambda\to0}
\frac{C(\Bin(m,p_\lambda))}{mp_\lambda}
=
\lim_{p\to0}
\frac{C(\Bin(m,p))}{mp},
\end{align*}
which completes the proof.
\end{proof}

\section{Proofs of the $\Del+\BSC$ Channel Results}\label{sec:proof_del_BSC}
In this section, we provide the proofs of the results presented in Section \ref{sec:del_bsc}.
\begin{proof}[Proof of Lemma \ref{lem:QPlusBSC}]
Since $X^n\to Y\to U$ forms a Markov chain, we have
$I(X^n;U\mid Y)=0.$
Therefore, using the chain rule for mutual information gives
\begin{align}
I(X^n;Y,U)
&=I(X^n;Y)+I(X^n;U\mid Y)=I(X^n;Y).
\label{eq:joint_mi_equals_y}
\end{align}
Applying the chain rule in the opposite order yields
\begin{align}
I(X^n;Y,U)
&=I(X^n;U)+I(X^n;Y\mid U).
\label{eq:chain_rule_u_first}
\end{align}
Combining \eqref{eq:joint_mi_equals_y} and
\eqref{eq:chain_rule_u_first}, we obtain
\begin{align}
I(X^n;U)
&=I(X^n;Y)-I(X^n;Y\mid U).
\label{eq:mi_decomposition}
\end{align}
Since $I(X^n;Y\mid U)\leq H(Y\mid U),$
it follows from \eqref{eq:mi_decomposition} that
\begin{align}
I(X^n;U)
&\geq I(X^n;Y)-H(Y\mid U).
\label{eq:mi_lower_by_cond_entropy}
\end{align}

We next upper bound $H(Y\mid U)$. Since the BSC does not change the sequence
length, $M$ is determined by both $Y$ and $U$. Therefore, we have
\begin{align}
H(Y\mid U)
&=H(Y\mid U,M).
\label{eq:add_length_condition}
\end{align}
Using the chain rule for conditional entropy, we have
\begin{align}
H(Y,U\mid M)
&=H(Y\mid M)+H(U\mid Y,M),
\label{eq:joint_entropy_y_first}
\end{align}
while
\begin{align}
H(Y,U\mid M)
&=H(U\mid M)+H(Y\mid U,M).
\label{eq:joint_entropy_u_first}
\end{align}
Combining \eqref{eq:add_length_condition},
\eqref{eq:joint_entropy_y_first}, and
\eqref{eq:joint_entropy_u_first}, we get
\begin{align}
H(Y\mid U)
&=H(Y\mid M)+H(U\mid Y,M)-H(U\mid M).
\label{eq:hy_given_u_identity}
\end{align}

Conditioned on $M=m$, the channel from $Y$ to $U$ consists of $m$ independent
uses of a $\operatorname{BSC}(\epsilon)$. Hence,
\begin{align}
H(U\mid Y,M=m)
&=m h_2(\epsilon).
\label{eq:bsc_cond_entropy_fixed_m}
\end{align}
Averaging \eqref{eq:bsc_cond_entropy_fixed_m} over $M$ gives
\begin{align}
H(U\mid Y,M)
&=\mathbb{E}[M]h_2(\epsilon).
\label{eq:bsc_cond_entropy_avg}
\end{align}
Since $M=\sum_{i=1}^n Q_i,$
where $Q_1,\ldots,Q_n$ are i.i.d. copies of $Q$, we have 
\begin{align}
\mathbb{E}[M]
&=n\mu_Q.
\label{eq:mean_output_length}
\end{align}
Therefore, \eqref{eq:bsc_cond_entropy_avg} yields
\begin{align}
H(U\mid Y,M)
&=n\mu_Q h_2(\epsilon).
\label{eq:bsc_cond_entropy_final}
\end{align}

For each $m\geq1$, define
\begin{align}
r_m:=\frac{H(Y\mid M=m)}{m}.
\label{eq:def_rm}
\end{align}
Conditioned on $M=m$, $Y$ is a binary vector of length $m$ and $U$ is obtained
by passing that vector through $m$ independent uses of a
$\operatorname{BSC}(\epsilon)$. Therefore, by Mrs.\ Gerber's Lemma \cite[Theorem 1]{wyner1973theorem}, we have
\begin{align}
H(U\mid M=m)
&\geq
m\,h_2\!\left(\epsilon\star h_2^{-1}(r_m)\right).
\label{eq:mgl_fixed_m}
\end{align}
Averaging \eqref{eq:mgl_fixed_m} over $M$ gives
\begin{align}
H(U\mid M)
&\geq
\sum_{m\geq1}\Pr(M=m)m
h_2\!\left(\epsilon\star h_2^{-1}(r_m)\right).
\label{eq:mgl_average}
\end{align}

From \cite[Lemma 2]{wyner1973theorem}, we know that $h_2\!\left(\epsilon\star h_2^{-1}(r)\right)$
is convex on $[0,1]$. Also, by
\eqref{eq:mean_output_length}, we have $\sum_{m\geq1}\frac{m\Pr(M=m)}{n\mu_Q}=1.$
Therefore, applying Jensen's inequality to \eqref{eq:mgl_average} with weights
$\frac{m\Pr(M=m)}{n\mu_Q}$ gives
\begin{align}
H(U\mid M)
&\geq
n\mu_Q
h_2\!\left(
\epsilon\star h_2^{-1}\!\left(
\frac{\sum_{m\geq1}\Pr(M=m)m r_m}{n\mu_Q}
\right)\right).
\label{eq:jensen_step}
\end{align}
By the definition of $r_m$ in \eqref{eq:def_rm}, we have
\begin{align}
\sum_{m\geq1}\Pr(M=m)m r_m
&=H(Y\mid M).
\label{eq:rm_average}
\end{align}
Using \eqref{eq:def_rbar} and \eqref{eq:rm_average} in
\eqref{eq:jensen_step}, we obtain
\begin{align}
H(U\mid M)
&\geq
n\mu_Q
h_2\!\left(
\epsilon\star h_2^{-1}(\bar r_n)
\right).
\label{eq:hu_given_m_lower}
\end{align}
Substituting \eqref{eq:def_rbar}, \eqref{eq:bsc_cond_entropy_final}, and
\eqref{eq:hu_given_m_lower} into \eqref{eq:hy_given_u_identity}, we get
\begin{align}
H(Y\mid U)
&\leq
n\mu_Q
\left[
\bar r_n+h_2(\epsilon)
-h_2\!\left(
\epsilon\star h_2^{-1}(\bar r_n)
\right)
\right].
\label{eq:hy_given_u_upper2}
\end{align}
Finally, substituting \eqref{eq:hy_given_u_upper2} into
\eqref{eq:mi_lower_by_cond_entropy} and dividing by $n$ completes the proof.
\end{proof}

\begin{lemma}
\label{lem:gepsNonincreasing}
Let $0\leq \epsilon\leq 1/2$ and define
\begin{align}
g_\epsilon(r)
:=
h_2\!\left(\epsilon\star h_2^{-1}(r)\right)-r,
\qquad r\in[0,1].
\label{eq:def_g}
\end{align}
 Then $g_\epsilon$ is nonincreasing on
$[0,1]$. 
\end{lemma}

\begin{proof}
By the chain rule,
\begin{align*}
g_\epsilon'(r)
&=(1-2\epsilon)
\frac{h_2'(\epsilon\star h_2^{-1}(r))}{h_2'(h_2^{-1}(r))}-1\\
&\leq(1-2\epsilon)-1=-2\epsilon\leq0,
\end{align*}
where the inequality follows from
$p\leq\epsilon\star p\leq1/2$ for $p
\in [0,1/2]$ and the fact
that $h_2'$ is nonnegative and decreasing
on $(0,1/2)$. 
\end{proof}

\begin{proof}[Proof of Lemma \ref{lem:HYMupperBound}]
Let $T$ denote the number of runs in $Y$, where a run
is a maximal consecutive sequence of identical bits.
We set $T=0$ when $M=0$.
Since $T$ is determined by $Y$, we have $H(T\mid M,Y)=0$, and hence, the chain rule gives
\begin{align}
H(Y\mid M)&=H(Y,T\mid M)\cr
&=H(T\mid M)+H(Y\mid M,T).
\label{eq:output_entropy_decomposition}
\end{align}

For $M=m\geq1$ and $T=t$, the output sequence
consists of $t$ runs with positive integer lengths
$k_1,\ldots,k_t$ satisfying
\begin{align}
k_1+\cdots+k_t=m,
\qquad k_i\geq1.
\label{eq:run_length_partition}
\end{align}
There are $\binom{m-1}{t-1}$ solutions to
\eqref{eq:run_length_partition}. Since the first
run can consist of either zeros or ones, the number
of possible output sequences is
$2\binom{m-1}{t-1}$. Consequently, we have
\begin{align}
H(Y\mid M=m,T=t)
&\leq 1+\log_2\binom{m-1}{t-1}
\nonumber\\
&\stackrel{(a)}{\leq} 1+\log_2\binom{m}{t}
\nonumber\\
&\stackrel{(b)}{\leq} 1+m h_2\!\left(\frac{t}{m}\right),
\label{eq:conditional_output_entropy_bound}
\end{align}
where $(a)$ follows from
$\frac{m}{t}\binom{m-1}{t-1}=\binom{m}{t}$, and $(b)$ follows from the standard
binomial entropy bound.

Averaging
\eqref{eq:conditional_output_entropy_bound}
over $M$ and $T$ yields
\begin{align}
H(Y\mid M,T)
&\leq
1
+\sum_{m\geq1}\sum_{t=1}^{m}
\Pr(M=m,T=t)\,
m h_2\!\left(\frac{t}{m}\right).
\label{eq:average_conditional_entropy}
\end{align}

For $m\geq1,\quad 1\leq t\leq m$, let
\begin{align*}
w_{m,t}:=
\frac{m\Pr(M=m,T=t)}{\mathbb{E}[M]}.
\end{align*}
Note that $\sum_{m\geq1}\sum_{t=1}^{m}w_{m,t}=1$.
We have
\begin{align}
\sum_{m\geq1}\sum_{t=1}^{m}&
\Pr(M=m,T=t)\,
m h_2\!\left(\frac{t}{m}\right)
\nonumber\\
&\quad=
\mathbb{E}[M]
\sum_{m\geq1}\sum_{t=1}^{m}
w_{m,t}h_2\!\left(\frac{t}{m}\right)
\nonumber\\
&\quad\stackrel{(a)}{\leq}
\mathbb{E}[M]\,
h_2\!\left(
\sum_{m\geq1}\sum_{t=1}^{m}
w_{m,t}\frac{t}{m}
\right)
\nonumber\\
&\quad=
\mathbb{E}[M]\,
h_2\!\left(
\frac{\mathbb{E}[T]}{\mathbb{E}[M]}
\right),
\label{eq:weighted_jensen_bound}
\end{align}
where $(a)$ follows from Jensen's inequality and the concavity of $h_2$.

Moreover, each output run contains symbols generated
by at least one input position. Thus $0\leq T\leq n$,
which implies
\begin{align}
H(T\mid M)
&\leq H(T)\leq\log_2(n+1).
\label{eq:run_count_entropy_bound}
\end{align}
Combining
\eqref{eq:output_entropy_decomposition},
\eqref{eq:average_conditional_entropy},
\eqref{eq:weighted_jensen_bound}, and
\eqref{eq:run_count_entropy_bound}, we obtain
\begin{align}
H(Y\mid M)
&\leq
\log_2(n+1)+1
+\mathbb{E}[M]\,
h_2\!\left(
\frac{\mathbb{E}[T]}{\mathbb{E}[M]}
\right).
\label{eq:output_entropy_finite_bound}
\end{align}


We next compute the asymptotic expected number of
output runs. Observe that the $Q$-repetition channel
can be represented as a cascade of an i.i.d.\
deletion channel with deletion probability $q_0=\Pr(Q=0)$
and a repetition channel that does not delete
any symbols. Specifically, let $Y'$ denote the
output of the deletion channel, and let each
symbol of $Y'$ be independently repeated a
positive number of times according to the
conditional distribution of $Q$ given $Q>0$.
The resulting output has the same distribution
as $Y$, the output of the $Q$-repetition channel
driven by $X^n$.
Since the second channel only repeats symbols
and never deletes them, it neither creates new
runs nor removes existing runs from $Y'$.
Consequently, the number of runs $T$ in $Y$
has the same distribution as the number of runs
in the output of an i.i.d.\ deletion channel
with deletion probability $q_0$. Therefore,
\begin{align}
\label{eqn:ExpTeqTp}
    \mathbb{E}[T]=\mathbb{E}[T'],
\end{align}
where $T'$ denotes the number of runs in $Y'$.

By \cite[Lemma 3.1]{mitzenmacher2009survey}, the surviving symbols
of a symmetric first-order Markov input form
a symmetric first-order Markov chain. In our
notation, its transition probability is
\begin{align*}
\beta
:=\Pr(Y'_{j+1}\neq Y'_j)
=
\frac{1-\alpha}
{1-q_0(2\alpha-1)}.
\label{eq:output_transition_probability}
\end{align*}
Let $M'=|Y'|$ denote the number of surviving
symbols. Since every run begins either with
the first surviving symbol or with a transition
between two consecutive surviving symbols,
we can write
\begin{align*}
T'
=
\mathbf{1}_{\{M'>0\}}
+
\sum_{j=1}^{M'-1}
\mathbf{1}_{\{Y'_{j+1}\neq Y'_j\}},
\end{align*}
where $\mathbf{1}_A$ denotes the indicator function of the event $A$.



By the strong law of large numbers for the deletion
process, we have $M'/n\to 1-q_0$, almost surely, and hence, $M'\to \infty$, almost surely as $n\to\infty$.
We can write
\begin{align*}
\lim_{n
\to \infty}\frac{T'}{n}
&=\lim_{n
\to \infty}\frac{\mathbf{1}_{\{M'>0\}}}{n}+\frac{1}{n}\sum_{j=1}^{M'-1}
\mathbf{1}_{\{Y'_{j+1}\neq Y'_j\}}\\
&=\lim_{n
\to \infty}\frac{M'-1}{n}\frac{1}{M'-1}\sum_{j=1}^{M'-1}
\mathbf{1}_{\{Y'_{j+1}\neq Y'_j\}}\\
&=(1-q_0)\lim_{n
\to \infty}\frac{1}{M'-1}\sum_{j=1}^{M'-1}
\mathbf{1}_{\{Y'_{j+1}\neq Y'_j\}}\\
&= (1-q_0)\beta
\qquad\text{almost surely},
\end{align*}
where the last equality follows from  the strong law of large numbers and the fact that  the random variables
$\mathbf{1}_{\{Y'_{j+1}\neq Y'_j\}}$ are i.i.d.\ Bernoulli$(\beta)$.
Since $0\leq T'/n\leq 1$ bounded,
almost sure convergence implies
\begin{align*}
\lim_{n\to\infty}\frac{\mathbb{E}[T']}{n}
=(1-q_0)\beta
=
\frac{(1-\alpha)(1-q_0)}
{1-q_0(2\alpha-1)}.
\end{align*}
Therefore, 
combining $\mathbb{E}[M]=n\mu_Q$ with
\eqref{eqn:ExpTeqTp}, we obtain
\begin{align*}
\lim_{n\to\infty}
\frac{\mathbb{E}[T]}{\mathbb{E}[M]}
&=
\frac{(1-\alpha)(1-q_0)}
{\mu_Q[1-q_0(2\alpha-1)]}.
\end{align*}
Finally, dividing \eqref{eq:output_entropy_finite_bound}
by $n\mu_Q$ and taking the limit superior complete the proof.
\end{proof}

\begin{proof}[Proof of Corollary \ref{cor:DelBSCLowerBound}]
First, consider the Poisson repetition channel with
$\lambda=1.79$. By \cite{mitzenmacher2006simple},
a symmetric first-order Markov input with
$\alpha=0.77$ achieves
\begin{align*}
\liminf_{n\to\infty}
\frac{I(X^n;Y)}{n\lambda}
\geq 0.1185.
\end{align*}
Since $\mu_Q=\lambda$ and $q_0=e^{-\lambda}$,
Lemma~\ref{lem:HYMupperBound} gives 
\begin{align*}
\limsup_{n\to\infty}
\frac{H(Y\mid M)}{n\lambda}
&\leq h_2(\rho_\lambda),
\end{align*}
where
\begin{align*}
\rho_\lambda=\frac{(1-\alpha)(1-e^{-\lambda})}
{\lambda[1-e^{-\lambda}(2\alpha-1)]}
\approx 0.117645<0.1177.
\end{align*}
Combining this with Lemma~\ref{lem:QPlusBSC}
and using the fact that $g_\epsilon$ is
nonincreasing, we obtain
\begin{align*}
C(&\Pois(1.79)+\BSC(\epsilon))
\nonumber\\
&\quad\geq
1.79\Big[0.1185-h_2(\epsilon)+h_2(\epsilon\star 0.1177)-h_2(0.1177)\Big]\\
&\quad\geq
1.79\Big[
h_2(\epsilon\star0.1177)
-h_2(\epsilon)-0.404214
\Big].
\end{align*}
Applying Corollary~\ref{cor:DelPoisBSC}
completes the proof of \eqref{eq:DelBSCPoisLowerBound1}.

Next, consider the deletion channel with deletion
probability $d_i$. In this case,
$q_0=d_i$ and $\mu_Q=1-d_i$.
By Lemma~\ref{lem:HYMupperBound}, for a symmetric
first-order Markov input with
$\Pr(X_{i+1}=X_i)=\alpha_i$, we obtain
\begin{align}
\limsup_{n\to\infty}
\frac{H(Y\mid M)}{n(1-d_i)}
&\leq
h_2\!\left(
\frac{1-\alpha_i}
{1-d_i(2\alpha_i-1)}
\right).
\end{align}
Combining this with Lemma~\ref{lem:QPlusBSC}
and the achievable rates $R_i$ reported in
Table~I of \cite{drinea2007improved} yields 
\begin{align*}
C(&\Del(d_i)+\BSC(\epsilon))
\nonumber\\
&\quad\geq
R_i-(1-d_i)h_2(\epsilon)
+
(1-d_i)\left[
h_2(\epsilon\star a_i)-h_2(a_i)
\right],
\end{align*}
where
\begin{align*}
a_i=\frac{1-\alpha_i}{1-d_i(2\alpha_i-1)},\text{\ \ and \  \ }
b_i=h_2(a_i)-\frac{R_i}{1-d_i}.
\end{align*}
Rearranging establishes \eqref{eq:DelBSCPoisLowerBound2}.
The coefficients in
Table~\ref{tab:DelBSC_thresholds}
are rounded conservatively to preserve
the inequality.
\end{proof}

\section{Conclusions}
\label{sec:Conclusion}

In this paper, we developed a channel-composition framework
for studying the capacities of repetition channels.
By introducing suitable repetition stages before or after
a given channel, we established capacity comparisons
between deletion, binomial-repetition, and Poisson-repetition
channels. In particular, we derived improved lower bounds
on the deletion-channel capacity in the high-deletion
regime, as well as new upper bounds on the capacity of
Poisson-repetition channels.


We also established an asymptotic equivalence between
the normalized capacities of binomial-repetition and
Poisson-repetition channels. As a consequence, we proved
that the limiting normalized deletion-channel capacity
equals the supremum of the normalized Poisson-repetition
capacities, which is approached as the Poisson parameter
tends to zero. This result addresses the question of
asymptotic tightness of the Poisson-based lower bound
raised in \cite{mitzenmacher2006simple}, establishing
that the bound is asymptotically tight in the
high-deletion regime.

Furthermore, we extended the framework to deletion
channels followed by a binary symmetric channel.
By combining repetition-channel capacity comparisons
with Mrs.\ Gerber's Lemma, we obtained improved
capacity lower bounds for channels with both deletion
and substitution errors. Numerical comparisons
demonstrated that the resulting bounds improve upon the
existing lower bounds and extend the range of
substitution probabilities for which positive lower
bounds can be established.

Overall, our results demonstrate that channel composition
provides a useful alternative to directly analyzing
the information rates of synchronization error channels.
Further investigation of other repetition distributions
and channel compositions may lead to additional capacity
bounds and asymptotic characterizations, including
for channels with different types of synchronization
and substitution errors.

\section*{Acknowledgment}
The authors acknowledge the use of ChatGPT,
developed by OpenAI, for assistance with
language editing, proofreading, and improving
the clarity and presentation of the manuscript,
as well as for assistance in developing the
deletion--BSC capacity bounds.

\bibliographystyle{IEEEtran}
\bibliography{reference}

\end{document}